# Green reconstruction of MIL-100 (Fe) in water for high crystallinity and enhanced guest encapsulation

Barbara E. Souza,[a] Annika F. Möslein,[a] Kirill Titov,[a] James D. Taylor,[b] Svemir Rudić,[b] and Jin-Chong Tan*[a]

**Abstract:** MIL-100 (Fe) is a highly porous metal-organic framework (MOF), considered as a promising carrier for drug delivery, and for gas separation and capture applications. However, this functional material suffers from elaborated and toxic synthesis that may hinder its biomedical use and large-scale production to afford commercial applications. Herein, we report a 'green' mechanochemical water immersion approach to yield highly crystalline MIL-100 (Fe) material. Subsequently, we have harnessed this strategy for facile fabrication of drug@MOF composite systems, comprising (guests) 5-fluorouracil, caffeine, or aspirin encapsulated in the pores of (host) MIL-100 (Fe). Inelastic neutron scattering was uniquely used to probe the guest-host interactions arising from pore confinement of the drug molecules, giving additional insights into the reconstruction mechanism. Our results pave the way to the 'green' production of MIL-type materials and bespoke guest-encapsulated composites by minimizing the use of toxic chemicals, whilst enhancing energy efficiency and material's life cycle central to biotechnological applications.

## Introduction

In the vast family of metal-organic frameworks (MOFs), the iron (III) carboxylate MOFs are promising given their biologically and environmentally favorable characteristics, large surface areas and high porosity. They offer numerous coordinatively unsaturated metal sites (CUS),[1] which not only potentializes their application as drug delivery systems (DDS), but also holds great promise as adsorbents and catalysts agents.[2-3] Specifically, MIL-100 (Fe) [MIL = Materials of Institute Lavoisier] exhibits a highly ordered structure with large pores (~3 nm), enabling the entrapment of large amounts of functional guests and drug molecules. As a functional material targeting biomedical applications, MIL-100 (Fe) features improved biocompatibility in contrast to a variety of MOFs, including its other metal counterparts, namely MIL-100 (Cr, Ni, Cu, and Co).[4]

Thus far, there have been several different reported approaches for the fabrication of MIL-100 (Fe). Since its discovery,[5] procedures have evolved from synthesis methods with long reaction times under harsh conditions (*i.e.* high temperature and high pressure coupled with the use of hydrofluoric acid (HF) and concentrated nitric acid ($HNO_3$)) to high pressure solvent-free procedures.[6-9] On the one hand, the use of toxic substances, such as mineralizing agents, to improve the crystallinity of MIL-100 (Fe) material could affect its applicability in the biomedical field.[10] On the other hand, improved crystallinity is crucial as it determines the material long-range periodicity and impacts its porosity and structural robustness. Today, one of the biggest challenges remains to optimize the synthetic procedures to yield highly crystalline MIL-100 (Fe) under mild conditions, i.e. avoiding the use of HF, toxic agents, high pressure and temperature. The harsh methods available not only jeopardize its potential bio-oriented applications, but also limit its scalability to enable commercial production. The ideal synthesis route should concomitantly allow the facile encapsulation of different (guest) drug molecules whilst preserving the structural properties and porosity of the (host) MIL-100 framework.

Herein, we present the use of a 'green' mechanochemical approach, which is free from the need to use HF or extreme processing conditions, to yield MIL-100 (Fe) at ambient conditions. We show that the crystallinity of the material can be significantly improved upon simple immersion in water (process termed: reconstruction). Notably, the reconstruction method is highly effective for the recovery of time degraded and mechanically amorphized samples, applicable as a material regeneration strategy. Additionally, the reconstruction step of MIL-100 (Fe) crystals can be tailored to enable the fabrication of guest@MIL-100 composite systems (*i.e.* guest@MIL-100_REC). By employing a variety of drug molecules (e.g. 5-fluorouracil, caffeine, and aspirin) as guests for confinement in the pores of MIL-100 (Fe), we study the vibrational dynamics of the drug@MOF systems *via* inelastic neutron scattering (INS at TOSCA spectrometer).[11-12] Analysis of the collective modes unravels details behind the reconstruction process, casting new light on the guest-host interactions underpinning the controlled drug release from MOF carriers.[13]

The guest molecules were chosen based on ongoing interest in their pharmaceutical applications, yet not fully realized due to existing drawbacks. 5-Fluorouracil (5-FU) is a long standing anti-cancer drug with an hydrophilic character, which restricts its passage through cell membranes without the aid of a DDS.[14] Caffeine (CAF) is widely applied as an active ingredient in

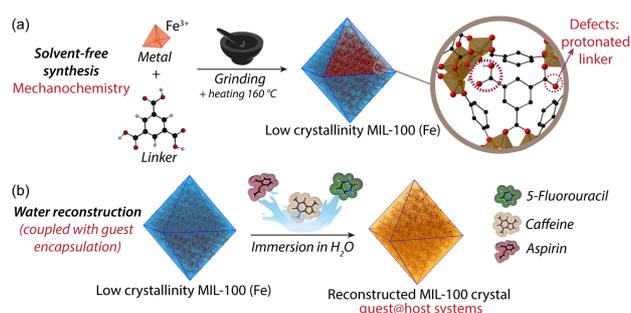

**Figure 1.** Schematic representation of the reconstruction method. (a) Mechanochemical approach applied to the synthesis of low crystallinity MIL-100 (Fe). (b) Reconstruction process to enhance material crystallinity and for the entrapment of different drug molecules within the MOF pores.

[a] B. E. Souza, A. F. Möslein, Dr. K. Titov, Prof. J. C. Tan
Multifunctional Materials & Composites (MMC) Lab, Department of Engineering Science, University of Oxford
Parks Road, Oxford OX1 3PJ, UK
E-mail: jin-chong.tan@eng.ox.ac.uk

[b] Dr. J. D. Taylor, Dr. S. Rudić
STFC Rutherford Appleton Laboratory
ISIS Neutron and Muon Source
Chilton, Didcot OX11 0QX, UK

Supporting information for this article is given via a link at the end of the document.

many cosmetic and pharmaceutical formulations. Increasing the loading percentage of caffeine within host carriers can hugely increase its bioavailability.[15] Aspirin (ASP) is commonly used as an analgesic and anti-inflammatory drug. However, many side effects (*e.g.* stomach bleeding and gastrointestinal ulcers) associated with its oral administration, can potentially be prevented by the use of a DDS.[16] A summary of the synthetic routes being employed in this study is illustrated in Figure 1, with full details given in the Supporting Information (SI).

## Results and Discussion

Figure 2 illustrates the changes observed in MIL-100 (Fe) as a function of time throughout the reconstruction process. We have discovered that upon immersion of the low-crystallinity sample (as-synthesized) into water, it is possible to significantly enhance the material crystallinity. The sample presented a distinctive colour change (Figure 2a) that was accompanied by the increase in its crystallinity, confirmed by the powder X-ray diffraction (PRXD) patterns (Figure 2b-c). The evolution in the material optical properties suggests the occurrence of microstructural changes in MIL-100 (Fe). A closer look at the diffraction data of as-synthesized MIL-100 (Fe) shows that the Bragg peaks below $2\theta = 5°$ exhibit either very low intensity or are completely absent from the pattern prior to the water reconstruction. Conversely, Bragg peaks within $2\theta = 10°$-$11.5°$ show a less pronounced change before and after reconstruction. The low relative intensity of the Bragg peaks at the smaller diffraction angles of $2\theta < 5°$ signifies a weak long-range ordering of the as-synthesized MIL-100 (Fe) framework. Because the intensity of the diffraction peaks is determined by periodic atomic arrangement averaged over the entire polycrystalline sample,[17] the overall increase in intensity observed in the PXRD patterns indicates the progressive increase of long-range ordering of the MIL-100 (Fe) crystals during reconstruction.

We further assessed the structural changes by analysing the relative peak intensity data as a function of sample immersion time (Figure 2b). The changing relative intensity of the (022):(357) planes, corresponding to the two most intense diffraction peaks, was calculated as the ratio between the peaks intensity. The results reveal a notable rise in intensity of the (022) plane (*i.e.* $2\theta = 4°$), versus a slower increase in intensity of the (357) plane (*i.e.* $2\theta = 11°$). The process can be divided into three stages. Firstly, during the first two days of immersion, the relative intensity of (022):(357) peaks were marginally unaffected. In the second stage (2-7 days), we observed a steep rise in the (022):(357) relative intensity from 0.57:1 to 1:0.27, indicating a major increase in long-range periodicity of MIL-100. Finally, the relative intensity ratio stabilized, with minimal changes from 7-15 days.

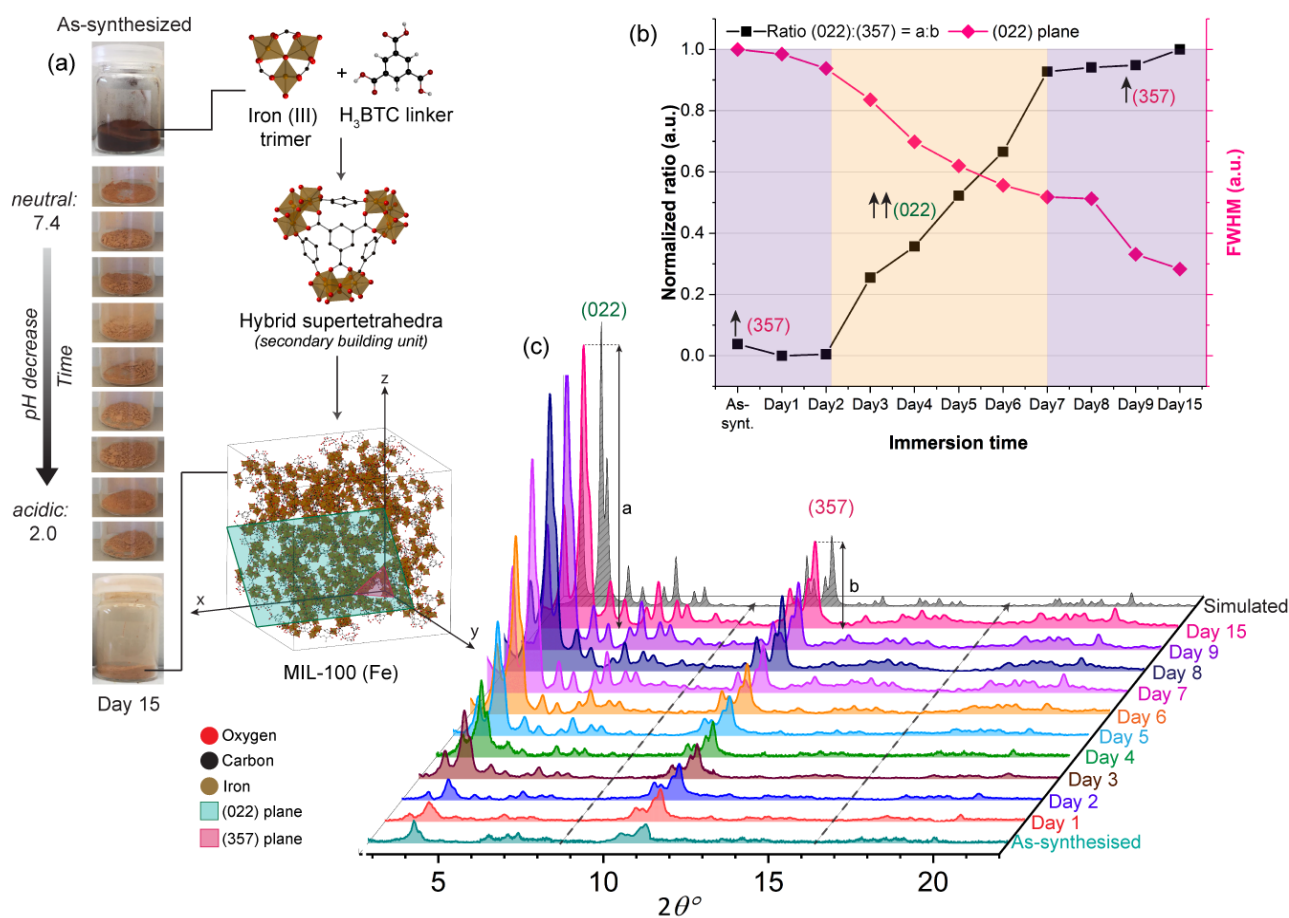

**Figure 2.** MIL-100 (Fe) reconstruction process. (a) Photographs showing evolution of the MIL-100 (Fe) samples collected at different immersion time intervals, and schematics of the associated chemical structure of MIL-100 (Fe) crystals. The images display the colour shift observed upon the increase of the crystallinity of MIL-100 (Fe) material. (b) Ratios of the changing relative intensity of the (022):(357) planes. The plots display a faster increase in peak intensity for small diffraction The plot also presents the FWHM of the peaks corresponding to the (022) plane, showing a progressive increase in the material crystallinity. (c) Evolution of the absolute intensity of the diffraction peaks throughout the reconstruction process, showing the marked improvement in crystallinity.

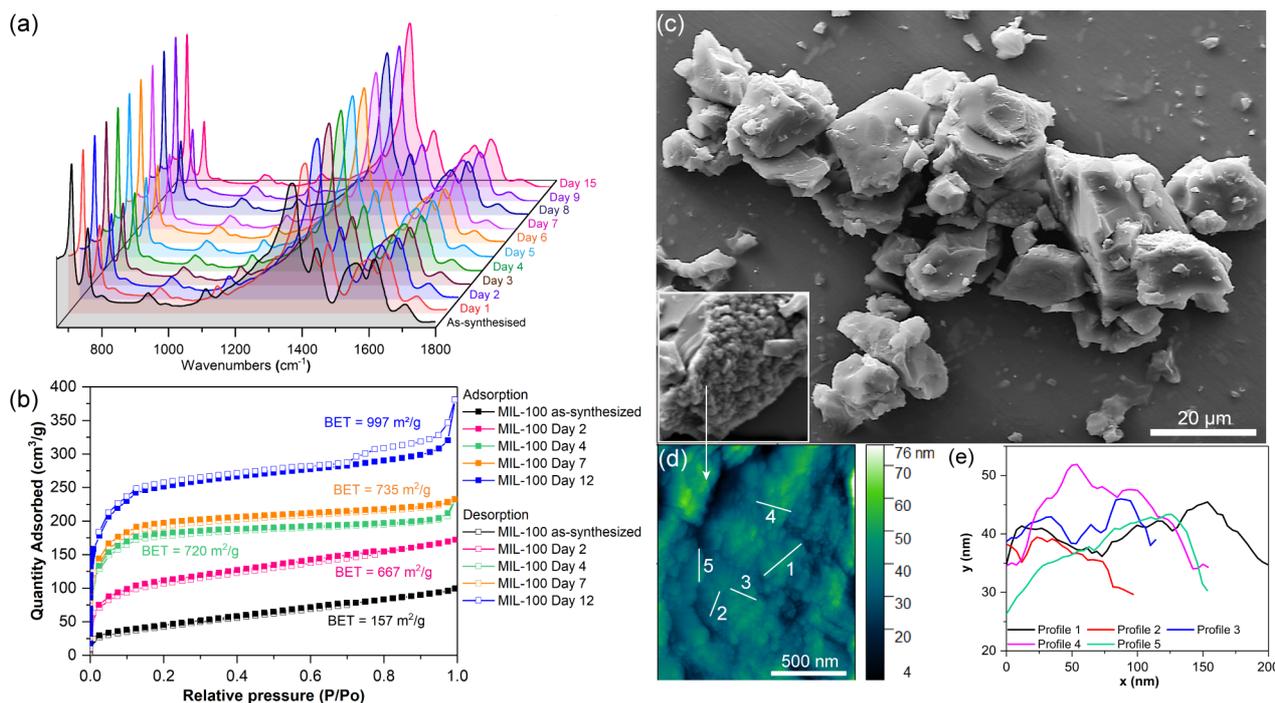

**Figure 3** (a) ATR-FTIR spectra of MIL-100 (Fe) samples collected at different immersion time intervals. (b) Nitrogen adsorption and desorption isotherms of MIL-100 (Fe) samples under reconstruction. Samples were activated at 150 °C under high vacuum for 12 hours prior to $N_2$ adsorption measurements at 77K. (c) SEM images of MIL-100 (Fe) samples. (d) AFM image displaying the surface morphology of MIL-100 (Fe) crystals accompanied by (e) the height profile of the marked regions, showing the formation of large particles by aggregation of nanosized particles.

angles below 10°. Notably, the change in the relative intensity of the peaks is accompanied by narrowing in FWHM (full width at half maximum) of the (022) peak, indicating increasing sample crystallinity. For comparison, the normalized PXRD patterns are presented in Figure S1 in SI.

Figure 3a shows the attenuated total reflectance Fourier transform infrared (ATR-FTIR) spectra where the typical vibrational bands of MIL-100 (Fe) were detected in all samples, confirming the chemical bonds integrity even in the low crystallinity material. From the ATR-FTIR spectra, two features can be highlighted. First, the sharpening of the band at ~1355 cm$^{-1}$, assigned to the stretching of the carboxylate groups present in the organic ligand (BTC = benzene-1,3,5-tricarboxylate), was observed (see Figure S2a). This change was quantified by the decrease in FWHM of this band (Figure S2b), indicating the increase in the structural symmetry of MIL-100 (Fe) samples.[18] Secondly, the band at ~1720 cm$^{-1}$, assigned to the stretching of carboxyl group present in the acidic form of the organic linker (*i.e.* H$_3$BTC = benzene-1,3,5-tricarboxylic acid),[8, 19] has shown a progressive decline in intensity, corresponding to the reduction in the spectral area of this vibrational peak (Figure S2b). These changes were accompanied by a large rise in the acidity of the solution in which MIL-100 (Fe) crystals were immersed (*i.e.* pH values fell from 7.4 to 2.0), as shown in Figure 2a, strongly indicating that a continuous deprotonation of the organic linker took place throughout the reconstruction process (days 1 to 15). A complete deprotonation of H$_3$BTC is required to produce a three-dimensional open framework with high crystallinity. The incomplete deprotonation of the organic ligand may obstruct the self-assembly of the secondary building units (Figure 2a), thereby resulting in partially formed (defective) mesocages. Similar results have been reported by Fernández-Bertrán *et al.*, who obtained partially formed frameworks of Zn(Imidazolate)$_2$ *via* a similar manual grinding process.[20] We propose that the presence of water, acting as a weak base to form hydronium ions (H$_3$O$^+$), will facilitate the deprotonation of H$_3$BTC and aid in the reconstruction of the defective MIL-100 (Fe) structure resultant from the grinding process.

To acquire a better understanding of the framework reconstruction process, we determined the Brunauer-Emmett-Teller (BET) surface areas of the samples using nitrogen sorption at 77 K. The BET results of four selected samples obtained from immersion times of 2, 4, 7 and 12 days are shown in Figure 3b and Table S2. We found a considerable increase (exceeding 500%) in surface area of the Day-12 sample compared with the as-synthesized MIL-100 (Fe). Evolution of the shape of isotherms at $P/P_o$ > ~0.7 demonstrates the reconstruction of mesocages in MIL-100 (Fe). From the as-synthesized to Day-12 samples, one can observe the progression from type II isotherm (non-porous materials) to type I, associated with microporous materials (*i.e.* pore size < 2 nm), and eventually reaching type IV, attributed to mesoporous cages (*i.e.* pore size = 2 - 50 nm).[21] As shown in Figure S3 in the SI, our reconstruction method offers a good trade-off between surface area and ease of synthesis, against other reported mechanochemical, solvothermal and solid-state strategies for synthesizing MIL-100 (Fe). For instance, in the work by Pilloni *et al.*,[22] in which a milling approach was used in combination with corrosive tetramethyl ammonium hydroxide, resulted in a material with comparable surface area to our reconstructed MIL-100 (Fe). The former additive can hinder the green production of MIL-100 (Fe) in terms of energy consumption and toxicity of reactant involved in the synthesis method.

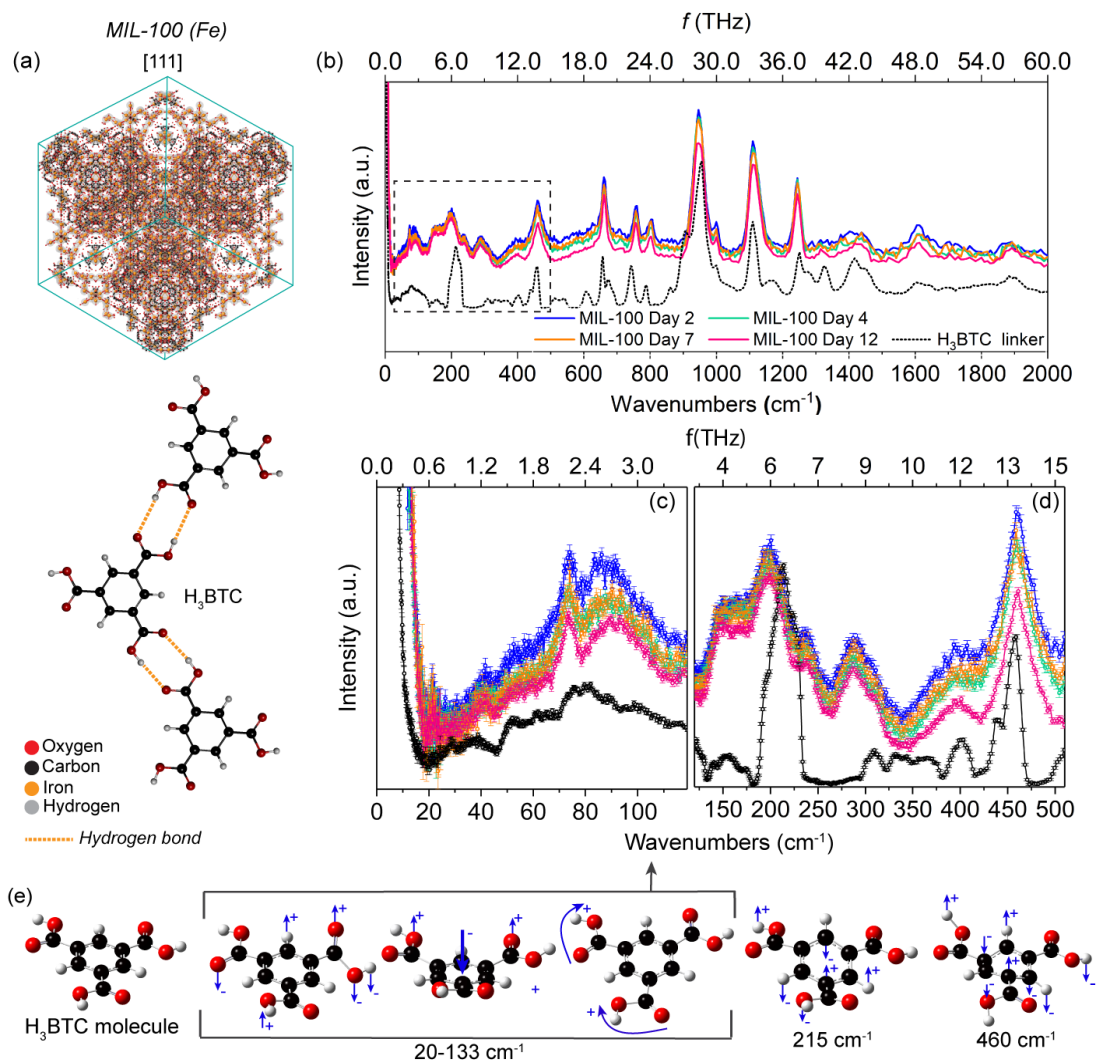

**Figure 4.** Inelastic neutron scattering measurements of reconstructed MIL-100 (Fe) samples. (a) Representation of MIL-100 (Fe) unit cell and H$_3$BTC molecules interacting *via* hydrogen bonds. (b) Inelastic neutron scattering (INS) spectra of MIL-100 (Fe) samples measured after different immersion times are presented against the spectrum of H$_3$BTC linker (scaled down to facilitate comparison to the MIL framework spectra). Closer look at the (c) low energy region (0-120 cm$^{-1}$), and (d) 145-510 cm$^{-1}$ range. (e) Schematic representation of H$_3$BTC molecular vibrations at different wavenumbers, as simulated by DFT.

Scanning electron microscopy (SEM) and atomic force microscopy (AFM) were used to characterize the morphology of the reconstructed samples (Figure 3c and Figures S4-S5). We observed particles in a large range of sizes, ranging from *ca.* 100s of nm to 10s of μm. The grinding process appears to favour the formation of aggregates of MIL-100 (Fe) crystals, culminating in larger particles (Figure 3c) constituted from densely packed nanoparticles (Figure 3c inset). AFM imaging (Figure 3d) revealed the granular nature of the particle surfaces, comprising nanocrystals < 50 nm as depicted in Figure 3e (see Figure S5 in the SI for further details).

Exceptionally, we have demonstrated the applicability of the reconstruction process to enable the recovery of time degraded samples, which were aged for 1.5 years (Figure S6-S7, SI). Likewise, we discovered that mechanically amorphized samples of MIL-100 (Fe) (Figure S8-S9, SI) can be repaired *via* water reconstruction; see SI for further discussion. This is an important finding, because it paves the way to the regeneration and recycling of MIL-100 materials, thereby addressing one of the main challenges preventing their industrial applications.[23-24]

Figure 4 displays the INS spectra of MIL-100 (Fe) samples collected after 2, 4, 7, and 12 days of immersion in water. Neutron scattering is a powerful spectroscopic method not subject to the optical selection rules. Unlike optical techniques such as infrared or Raman, all transitions are active in INS spectra.[25] The neutron is a highly sensitive probe to measure local changes in vibrational modes, especially low energy phonons or terahertz (THz) vibrations.[26] These vibrations are collective modes associated with the lattice dynamics of the framework structure and their analysis *via* INS spectroscopy can provide further insights into how the structure of MIL-100 (Fe) is rebuilt during the reconstruction process. Due to the difficulty in isolating the origin of each THz vibration in a complex system like MIL-100 (Fe), we have collected the INS spectrum of the H$_3$BTC linker (Figure 4a) and correlated it with density functional theory (DFT) calculations.[27] The complete comparison between experimental and theoretical spectra of H$_3$BTC accompanied by description of the vibrational modes can be found in Figure S10 and Table S3. As displayed in Figure 4a, the H$_3$BTC molecules may establish hydrogen bonding interactions via –COOH groups. By contrasting the linker and framework spectra, we identified the modes that are predominantly influenced by the BTC linker within the framework structure of MIL-100 (Fe). A good agreement between H$_3$BTC and MIL-100 (Fe) spectra was observed, both in the low and high energy regions (see Figure 4b and Figure S11, SI).

It can be seen in Figures 4b-d that the overall MIL-100 (Fe) INS spectral intensity declines with a higher sample immersion time. We can attribute this decay to two factors. First, in the INS spectrum generated on TOSCA the intensity at low frequency (i.e. low momentum transfer (Q)) is proportional to the mean square displacement of the atoms from their equilibrium position. At higher frequency, however, the intensity is suppressed by the Debye–Waller factor.[28] On this basis, the decline in the scattering intensity in the low energy region, below 9 THz (< 300 cm$^{-1}$), with the increase of immersion time suggests the reduction of atomic motions. As the periodicity of the MIL-100 (Fe) samples improves, we reasoned that stronger constraints are imposed onto the atomic movements in the framework. This result is substantiated by the notable rise in the thermal stability of the reconstructed framework via thermogravimetric analysis (TGA, see Figure S12), where an ~48% increase in the initial decomposition temperature was recorded (see Table S4 for more details).

Another factor that contributes to the overall spectrum intensity is the amount of hydrogen atoms present in the sample, due to the large incoherent neutron cross-section of hydrogen atoms, the largest amongst all known elements.[29] We can therefore correlate the observed decrease in spectral intensity with the deprotonation of the organic linker, previously observed in the ATR-FTIR measurements (Figure S2b). As shown in the inset of Figure S11, the acidic form of the organic ligand (i.e. H$_3$BTC) has twice as many hydrogen atoms than its deprotonated counterpart (i.e. BTC). In the low energy region, H$_3$BTC vibrations are dominated by the out-of-plane bending of -COOH, including 'trampoline-like' motions,[30] at 0.6-4.0 THz (20-133 cm$^{-1}$), out-of-plane -CO bending at ~6.4 THz (215 cm$^{-1}$), and ring deformation accompanied by -OH bending at ~13.8 THz (460 cm$^{-1}$), as illustrated in Figure 4e.[27] These vibrational modes, involving the large displacement of hydrogen atoms, show a decline in intensity upon the increase of sample immersion time, thus, the INS results support the notion that the reconstruction of MIL-100 (Fe) mesocages is controlled by the deprotonation of the organic linkers.

We have discovered that this approach can additionally be applied to fabricate highly loaded guest@host systems (termed: guest@MIL-100_REC) leading to the encapsulation of different drug molecules (guest = 5-FU, CAF, and ASP) when MIL-100 (Fe) cages are being reconstructed (experimental details are presented in the SI). SEM images of guest@MIL-100 systems reveal that the drug-loaded samples produced via the reconstruction encapsulation approach have a similar morphology to the reconstructed MIL-100 (Fe) samples (Figure S13).

Figures 5a-c show the PXRD patterns of drug@MIL-100 composite systems. The relative intensity of (022):(357) planes and the FWHM of the (022) corresponding peak of the drug@MIL-100 systems were calculated and contrasted to assess the effect that the confinement of the different drug molecules have on the crystallinity of the host material. Through the reconstruction process, 5-FU@MIL-100_REC with high crystallinity was successfully obtained. Generally, we can see that the relative peak intensity in 5-FU@MIL-100_REC (i.e. 1:0.30) agrees well with pristine MIL-100 (Fe) (1:0.24) (Figure 5a). The FWHM in Figure S14 shows that the confinement of 5-FU within MIL-100 (Fe) pores did not affect the periodicity and crystallinity of the resultant framework. Similar effect was observed in CAF@MIL-100_REC, for which the relative peak intensity (i.e. 1:0.37) and the FWHM of (022) peak were slightly larger than in 5-FU loaded counterpart, but still in close agreement with MIL-100 (Fe) pristine material (Figures 5a and S14, SI). The fabrication of ASP@MIL-100_REC was correspondingly achieved, however, the periodicity of the material was reduced when contrasted to the other drug@MIL-100 systems. After the encapsulation of aspirin, the relative intensity of diffraction peaks (i.e. 0.9:1) was dissimilar to the value of MIL-100 (Fe) and a large broadening of the (022) peak was detected (Figure S14, SI). These results indicate that aspirin, in some degree, interferes with the establishment of long-range ordering of the host framework. This is because of the competition between aspirin and the H$_3$BTC linker molecules for coordination to the iron cations, leading to formation of an amorphous violet aspirin-iron complex known as tetraaquosalicylatroiron (III) (Figure S15, SI).

Figures 5d-f display the INS spectra of the drug@MIL-100 composite systems. The full spectra up to ~60 THz (2000 cm$^{-1}$) are shown in Figures S16-S19. Scrutiny of the THz vibrations can provide insights on the guest-host interactions and reveal how the encapsulation of the different drug molecules affects the long range dynamics of the host framework structure.[31] The spectra of the pure guest molecules were also collected and used for the assignment of the drug peaks in drug@MIL-100. Analysis of the INS spectra shows that in all guest@host systems considered, the distinctive vibrational peaks of guest molecules were detected. This is akin to what was also observed in the high energy region, in which the ATR-FTIR spectra of the drug@MOF systems (Figure S20, SI) also display the clear presence of the drug molecules. In order to correlate the influence of the guest:host ratio on the drug@MIL-100 spectra, the level of guest encapsulation was evaluated by BET surface area (Figure S21, SI) and quantified by TGA (Figure S22, SI). Overall, remarkably high drug loadings of 35.5 wt.% (0.6 g/g MOF), 64.7 wt.% (1.8 g/g MOF), and 70 wt.% (2.4 g/g MOF), were achieved for 5-FU@MIL-100_REC, CAF@MIL-100_REC, and ASP@MIL-100_REC, respectively.

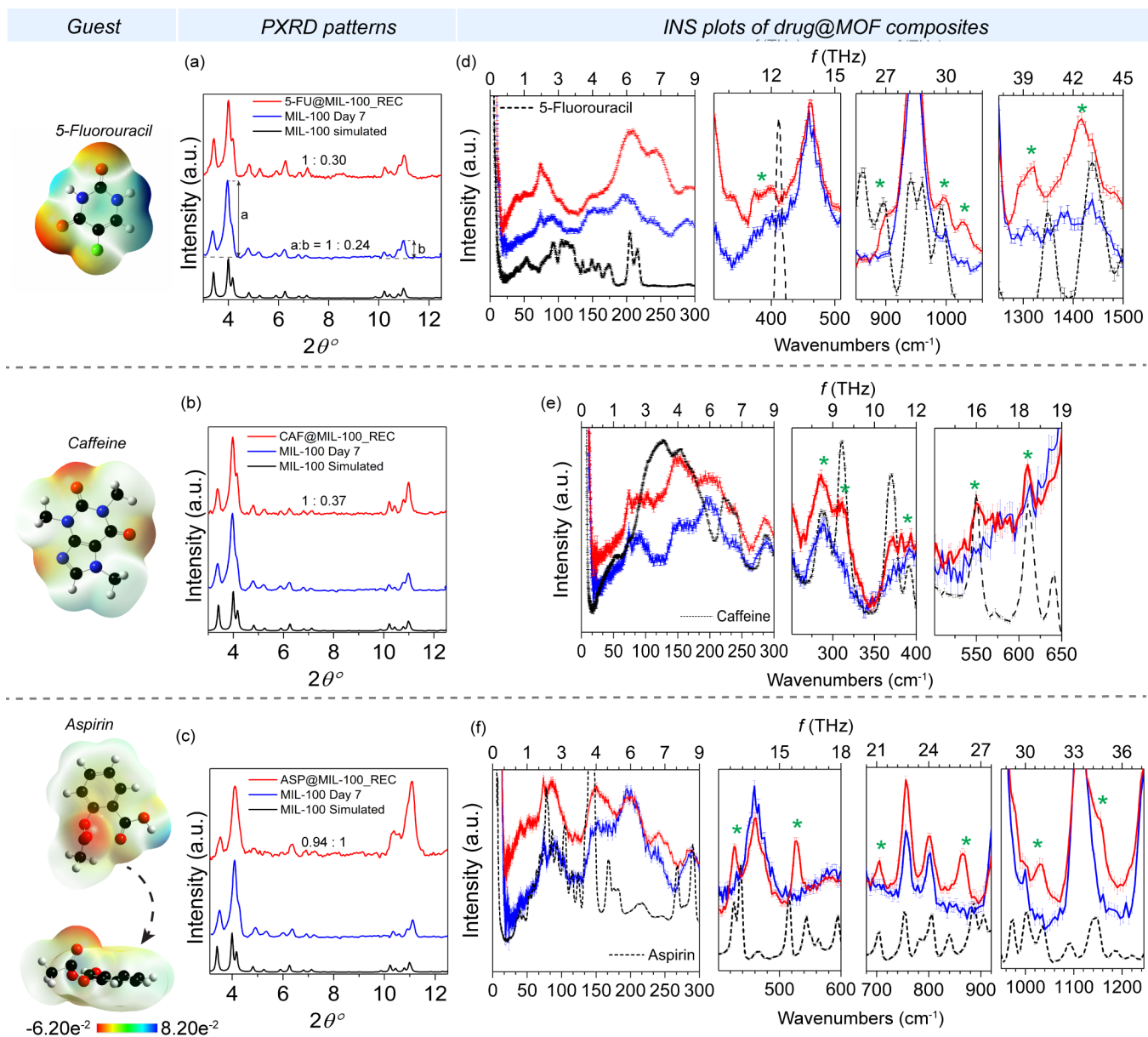

**Figure 5.** Powder X-ray diffraction patterns and inelastic neutron scattering (INS) spectra of drug@MIL-100 (Fe) systems. The diffraction patterns are presented for assessment of the materials crystallinity as (a) 5-FU@MIL-100_REC, (b) CAF@MIL-100_REC, and (c) ASP@MIL-100_REC. INS spectra of (d) 5-FU@MIL-100_REC, (e) CAF@MIL-100_REC, and (f) ASP@MIL-100_REC, with closer look at the drug peaks present in the drug@MIL-100 systems, highlighted with asterisks. The guest drug molecules and their respective electrostatic potential maps (ESP) were presented, for which aspirin is shown in two different angles for better visualization. Colour code: O in red, C in black, H in grey, N in navy blue, F in green.

The BET surface areas of the samples were found to be greatly reduced, confirming successful guest encapsulation. In Table S5, we compared the guest loading herein attained against previously reported results. In the work by Cunha et al.[32] a theoretical approach has been used to estimate the maximum loading of caffeine molecules in MIL-100 (Fe) mesocages. A theoretical loading capacity of 65.8 wt.% was established when full accessibility of both small and large cages (i.e. 25 and 29 Å in diameter) was considered. However, using the conventional impregnation technique (i.e. immersion of the host MOF into a saturated drug solution) an experimental loading of only 49.5 wt.% was attained by Cunha and co-workers. The difference between the theoretical loading capacity and the experimental loading were attributed to the non-accessibility of caffeine molecules (7.6 × 6.1 Å) into the small cages of MIL-100 (Fe) due to the reduced size of its pentagonal window aperture (~4.5 × 5.5 Å). Conversely, higher experimental loadings were achieved using our water reconstruction methodology (i.e. 64.7 wt.%). As illustrated in Figure 6, instead of having to overcome the physical obstacle imposed by the narrow pore windows, the guest drug molecules get encapsulated by the formation of the MIL-100 (Fe) cages around them. This improves the occupancy of the small and large cages allowing us to harness the full hosting potential of MIL-100 (Fe). The possibility of confinement of guest molecules in the small cages of MIL-100 (Fe) opens new frontiers to tuning the prolonged release of therapeutic molecules (avoiding burst effect[13] *via* the slow decomposition of pore cages) and to yield permanent encapsulation of bulky guests, such as fluorescent dyes for photoluminescent devices.[33]

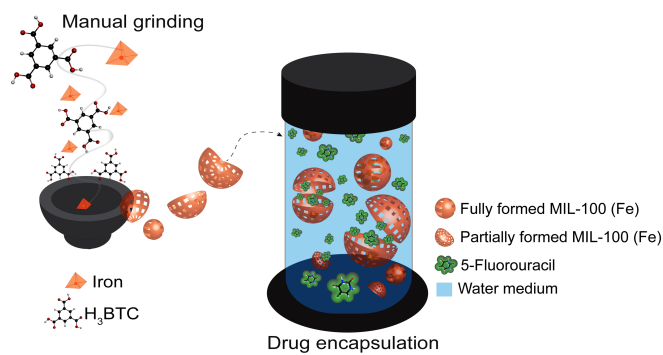

**Figure 6.** Schematic representation of the guest encapsulation mechanism during the reconstruction process immersed in water. The formation of the MOF cages around the drug molecules allows the achievement of high guest loadings.

Further insights into the dynamics of the guest-host interaction can be drawn by analyzing the INS spectra. The relatively higher intensity observed in the spectra of the guest@MIL-100_REC, specifically in the 0-9 THz (0-300 cm$^{-1}$) range, can be associated with a combination of scattering coming from the framework and the drug molecules. As seen in Figures 5d-f, the spectra of the drug molecules exhibit a very high scattering intensity in the low energy region, meaning that even low amounts of encapsulated guest can have a large effect on the vibrational spectra of the drug@MIL-100 system.

Vibrational frequency shifts were observed in the guest vibrational modes present in the drug@MIL-100 spectra (marked in Figures 5d-f by asterisks). Such shifts indicate the existence of constraints to free motions of the drug molecule. For example, the effect of the guest confinement is evident on the 5-FU@MIL-100_REC spectrum, in which the very strong 5-FU vibration at ~12.3 THz (~410 cm$^{-1}$) associated with the in-plane bending of OCNCO is highly suppressed (Figure S23).[34] Based on the electrostatic potential surface map (ESP) depicted in Figure 5, 5-FU molecules will bind to the CUS *via* one of their oxygen atoms through Fe-O coordination. The same scenario is repeated for caffeine and aspirin molecules, which as shown in the ESP maps in Figure 5, can also establish Fe-O coordination to the host MIL-100 (Fe) CUS. Similarly, the suppression and shift of the caffeine modes at ~10.8 THz (~360 cm$^{-1}$) related to C=O bending and the shift of aspirin modes at ~12.6 THz (~421 cm$^{-1}$), ~16.0 THz (~533 cm$^{-1}$) related to the C=O bending, substantiates the formation of Fe-O coordination between caffeine and aspirin to the CUS. Shift of the Aspirin modes at ~25.6 THz (~855 cm$^{-1}$) and ~31.2 THz (~1040 cm$^{-1}$), associated with benzene ring deformation suggests that the drug molecules simultaneously form π-π interactions with the organic linker and between drug molecules within MIL-100 (Fe) cages. In fact, comparing the size of MIL-100 (Fe) mesocages (25 and 29 Å in diameter),[5] the chosen drug molecules are considerably small (see Figure S24) meaning that the interactions between multiple drug molecules (*e.g.* dimers formation) are expected inside the same pore. The suppression and shifts of high intensity drug molecule modes indicate the strong guest-host interaction yielded by the employment of the reconstruction encapsulation method.

## Conclusions

In summary, we have demonstrated the use of a low cost, green mechanochemical method applied to the facile fabrication of highly crystalline MIL-100 (Fe) material, where framework reconstruction in water leads to improved thermal stability. This approach can be used to achieve guest@MIL-100 systems with a high loading of the encapsulated guest molecules. High-resolution inelastic neutron scattering spectroscopy aided the understanding of the guest-host interactions through unravelling the vibrational dynamics of the MIL-100 (Fe) phase and its guest-encapsulated composites. This work provides new approaches towards the eco-friendly and scalable synthesis of mesoporous MIL-100 (Fe) and guest@MIL-100 systems, avoiding the use of highly toxic agents that are limiting the promising biomedical applications of MIL-100 (Fe). The simple approach demonstrated here could be applied to the large-scale synthesis of mesoporous MIL-100 materials for future commercialization.

## Associated content

**Supporting Information (SI)**

Synthetic procedures in detail, DFT calculations, materials characterization (TGA, SEM, AFM, and PXRD), and drug encapsulation studies.

## Author information

Corresponding Author
*jin-chong.tan@eng.ox.ac.uk

**Funding**

This work was funded by the Minas Gerais Research Foundation (FAPEMIG), the European Research Council (ERC) and the Engineering and Physical Sciences Research Council (EPSRC).

## Acknowledgements

B. E. S. thanks the Minas Gerais Research Foundation (FAPEMIG CNPJ n21.949.888/0001-83) for a DPhil scholarship award. J. C. T. thanks the EPSRC Grant No. EP/N014960/1 and ERC Consolidator Grant under the grant agreement 771575 (PROMOFS) for funding. We acknowledge the ISIS Neutron and Muon Source for the award of beamtime no. RB1910059 during which the INS experiments were performed on the TOSCA spectrometer. We are grateful to the Research Complex at Harwell (RCaH) for the provision of advanced materials characterization facilities.

# Supporting Information

# Green reconstruction of MIL-100 (Fe) in water for high crystallinity and enhanced guest encapsulation

Barbara E. Souza,[a] Annika F. Möslein,[a] Kirill Titov,[a] James D. Taylor,[b] Svemir Rudić,[b] and Jin-Chong Tan[a]*

[a] Multifunctional Materials & Composites (MMC) Lab, Department of Engineering Science, University of Oxford, Parks Road, Oxford OX1 3PJ, UK

[b] STFC Rutherford Appleton Laboratory, ISIS Neutron and Muon Source, Chilton, Didcot OX11 0QX, UK

*J.C. Tan, email: jin-chong.tan@eng.ox.ac.uk

## Table of Contents

# 1. Materials synthesis

## 1.1. Mechanochemistry of MIL-100 and crystallinity reconstruction

The MIL-100 (Fe) MOF material was synthesized *via* a manual grinding-annealing process. Fe(NO$_3$)·9H$_2$O [iron(III) nitrate 9-hydrate] (3 mmol) with H$_3$BTC [benzene-1,3,5-tricarboxilic acid] (2 mmol) were combined in the agate mortar and manually ground for 10 min. The resulting material was heated in the oven at 160 °C for 4 h to complete the annealing process. The product was washed by centrifugation (8000 rpm for 10 min) with methanol and deionized water to remove any unreacted components. Pristine MIL-100 (Fe) was then dried at room temperature and activated under vacuum at 150 °C for 12 h.

To perform the reconstruction and crystallinity studies, MIL-100 (Fe) particles were immersed into deionized water and kept under stirring at room temperature. After different immersion times, the samples were separated by centrifugation (8000 rpm for 10 min) and the supernatant was collected to measure the pH values of the reconstruction solution. Finally, the recovered MIL-100 (Fe) samples were dried at room temperature and re-activated under vacuum at 150 °C for 12 h. The reconstruction process was repeated to ensure reproducibility of results acquired with reconstruction and the product obtained *via* the manual grinding method.

## 1.2. Drug@MIL-100 composite systems preparation

The drug@MOF systems were synthesized as follows. Encapsulation during the reconstruction process was performed by immersion of pre-activated MIL-100 (Fe) samples in a saturated aqueous drug solution (*i.e.* 12 mg/ml for 5-FU, 20 mg/ml for caffeine, and 3.3 mg/ml for aspirin) under continuous stirring for different time periods at room temperature. The drug-loaded drug@MIL-100_REC particles were separated by centrifugation (8000 rpm for 10 min) and then activated at 130 °C for 12 h in a vacuum oven.

**Table S1**: Samples description and details

|  | Samples | Synthesis method | Details |
|---|---|---|---|
| *Reconstruction studies* | MIL-100 (Fe) | Manual grinding | - |
|  | MIL-100 (Fe) Day x [a] |  | Samples collected after different x immersion times |
| *Encapsulation during reconstruction* | 5-FU@MIL-100_REC | Manual grinding (reconstruction encapsulation) | Guest: 5-Fluorouracil |
|  | CAF@MIL-100_REC |  | Guest: Caffeine |
|  | ASP@MIL-100_REC |  | Guest: Aspirin |

[a] x: immersion time (e.g. 1, 2, 3…15 days)

## 2. Materials characterization

### 2.1. Inelastic neutron scattering

Inelastic neutron scattering (INS) measurements were performed using the TOSCA[1] spectrometer at the ISIS Pulsed Neutron and Muon Source, Rutherford Appleton Laboratory (Chilton, UK). The high-resolution ($\Delta E/E$~1.25%) and broadband (0-4000 cm$^{-1}$) spectra of each sample (~1 g) were acquired at ~10 K.

TOSCA is an indirect geometry time-of-flight spectrometer where a pulsed, polychromatic beam of neutrons collides with the sample at a distance of ~17 m from the source. The neutrons scattered from the sample were Bragg reflected by a pyrolytic graphite analyzer, while higher-order reflections beyond (002) were blocked by a cooled ($T$ < 30 K) Beryllium filter in order to define the final energy. Neutrons with final energy of ~32 cm$^{-1}$ were passed towards the detector array composed by thirteen $^3$He tubes with effective length of 250 mm. Five banks were located in forward direction (scattering angle ~45°) and five in backwards direction (~135°). The use of a low final energy translated into a direct relationship between energy transfer ($E_T$, cm$^{-1}$) and momentum transfer ($Q$, Å$^{-1}$) such that $E_T \approx 16Q^2$. Energy transfer and spectral intensity, $i.e.$ $S(Q, \omega)$, were then obtained using the Mantid software.[2] Each sample was wrapped in 4 cm × 4.6 cm aluminium sachet and placed into a 2.0 mm spaced flat aluminium cell, which was sealed with indium wire. Sample preparation and cell loading into the cell took place in a glovebox to avoid moisture uptake by the sample. To reduce the effect of the Debye-Waller factor on the experimental spectral intensity and allow comparison with the theoretical spectra, the sample cell was cooled to ~10 K by a closed cycle refrigerator (CCR). The INS spectra were collected under vacuum over a duration of 4-6 hours.

The neutron guide upgrade of the TOSCA spectrometer, completed in 2017, has increased the neutron flux at the sample position by as much as 82 times. This upgrade improves the performance through faster measurements and by reducing the required sample mass.[3]

### 2.2. Powder X-ray diffraction

The powder samples were analyzed by powder X-ray diffraction (PXRD) using the Rigaku MiniFlex diffractometer with a Cu K$_\alpha$ source (1.541 Å). Diffraction data was collected from 3° to 13°, using a 0.02° step size and 0.1° min$^{-1}$ step speed. The patterns were then normalized with respect to the most intense peak in the pattern [0-1] used for the relative intensity of selected peaks. The raw PXRD data, however, were used for the calculation of FWHM peak broadening/sharpening.

### 2.3. Thermogravimetric analysis

Thermogravimetric analysis (TGA) was performed using TGA-Q50 (TA instruments). Approximately 4 mg of each sample was placed in a platinum pan (maximum volume 50 µL) and heated from 50 °C to 700 °C with a heating rate of 10 °C min$^{-1}$. The measurements for MIL-100 (Fe) samples were performed under a dry nitrogen flow of 40 mL min$^{-1}$. The measurements for drug@MIL-100 (Fe) samples were conducted under an air flow of 40 mL min$^{-1}$ to guarantee complete decomposition of the guest molecules and accurate guest loading determination.

### 2.4. Attenuated total reflectance Fourier transform infrared spectroscopy

Attenuated total reflectance Fourier transform infrared spectroscopy (ATR-FTIR) spectra were acquired at room temperature with a Nicolet iS10 FTIR spectrometer with an ATR sample holder. The spectra were collected in the range of 650-4000 cm$^{-1}$ with a resolution of 0.5 cm$^{-1}$ and normalized in respect to the most intense vibrational peak to facilitate comparison across the different samples under study.

### 2.5. Scanning electron microscopy and atomic force microscopy

Analyses of the morphology and particle size determination were carried out by scanning electron microscopy (SEM) and atomic force microscopy (AFM). SEM images were obtained using Carl Zeiss EVO LS15 at 15 keV under high vacuum. Atomic force microscopy (AFM) was performed using the Veeco Dimension 3100 AFM equipped with an in-line optical zoom microscope with colour CCD camera for precise placement of probe onto the sample. The microscope was operated under the tapping mode, equipped with a Tap300G silicon probe with resonance frequency of 30 kHz, spring constant of 40 Nm$^{-1}$, and tip radius < 10 nm.

### 2.6. Specific surface area measurements

The Brunauer-Emmett-Teller (BET) specific surface areas of samples were determined from nitrogen adsorption-desorption isotherms at 77 K, measured with Quantachrome Nova 1200. The isotherms were obtained using a ⌀9 mm sample cell containing 60-100 mg of samples under study. The degassing temperature was 150 °C during sample activation under vacuum.

### 2.7. Density functional theory calculations

Density functional theory (DFT) calculations to determine the theoretical vibrational spectrum of H$_3$BTC was performed using the Gaussian software.[4] The vibrational calculation was carried out at the B3LYP level of theory and 6-31G basis set. We used the DFT output file of this study to generate the INS spectrum using the Mantid software[2] through the AbINS extension.[5] During the spectrum generation a total cross section was considered with a quantum order events number of 1.

**2.8. Calculation of PXRD peaks height and full width at half maximum of ATR-FTIR peaks**

The Integrate Gadget in OriginPro software was used to perform the numerical integration on the PXRD patterns and determine the full width at half maximum (FWHM) of two of the most intense diffraction peaks in the MIL-100 (Fe) samples (i.e. $2\theta$ = 4° and 11°) that correspond to the (022) and (357) planes. The range of data was selected to include the peaks of the diffraction pattern of interest, using the (horizontal) diffraction angle axis as the baseline. The same approach was used for the obtaining the FWHM of ATR-FTIR vibrational bands. To facilitate the comparison between the effect of the reconstruction process, the ratio between the peak heights [(022):(357)] was taken and the FWHM values were normalized against the largest value presented within a set of samples.

**2.9. Acidity measurements of immersion solutions**

The pH values of the immersion solutions were determined using a Fisherbrand pH indicator paper stick that was compared against a pH scale and across all the samples in the study.

**2.10. Proton acceptor/donor sites and electrostatic potential surface map determination**

The proton acceptor/donor sites present in the guest drug molecules (*e.g.* 5-fluorouracil, caffeine, and aspirin) were determined using BIOVIA Discovery Studio. The electrostatic potential (ESP) surface maps of drug molecules were generated using GaussView. During the map generation, the electron density was calculated from a Total SCF density with isovalue of 0.000400 electrons per unit volume (au$^3$).

**2.11. Fabrication of pellets**
Pellets were prepared on a manual hydrostatic press (Specac) with a die diameter of 13 mm and under a constant axial force that varied tom 0.5 to 10 ton. Each pellet was produced using 175 mg of the MOF material.

## 3. Results and Discussion

### 3.1. Reconstruction of as-synthesized MIL-100 (Fe)

The effect of the reconstruction process has been monitored *via* analysis of the PXRD data. The evolution of the relative peak intensity [*i.e.* (022):(357) ratio] and of the change in FWHM at low diffraction angles (2θ = 4° and 11°) has been monitored as a function of the sample immersion time.

The Scherrer law allows one to determine the size of the crystalline domains $D$:[6]

$$D = \frac{K\lambda}{\Delta \cos\theta}$$

where $\lambda$ is the wavelength, $\Delta$ = FWHM, $K$ *is* a constant, and $\theta$ is the diffraction angle of the corresponding diffraction peak.

As the samples might present different particle sizes (from 100s of micrometers to 10s of nanometers) and different packing configurations, the best representation to establish the cross comparison of such samples was chosen as the "crystallinity" where:

$$Crystallinity = \frac{D_{\text{initial}}}{D_{\text{x}}}$$

With *x* referring to the different immersion times. Then:

$$Crystallinity = \frac{\frac{K\lambda}{\Delta_{\text{initial}}\cos\theta}}{\frac{K\lambda}{\Delta_{\text{x}}\cos\theta}} = \frac{\Delta_{\text{x}}}{\Delta_{\text{initial}}}$$

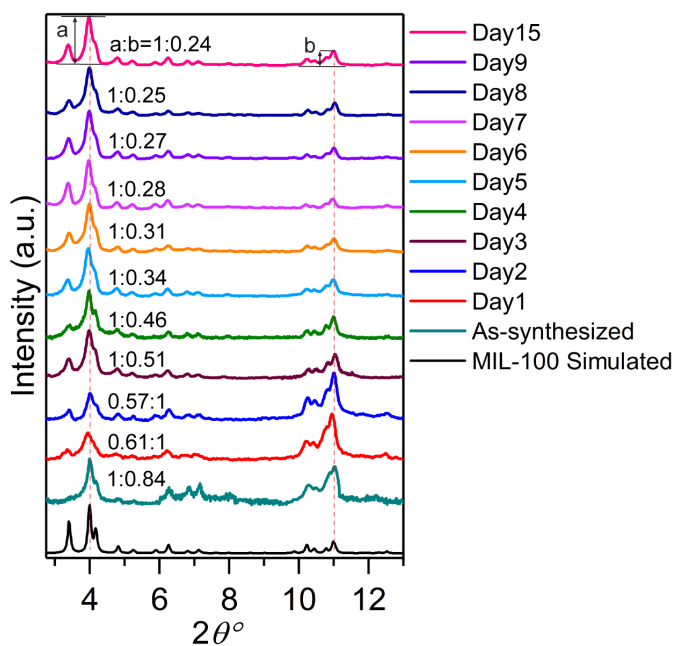

**Figure S1.** Normalized PXRD patterns of MIL-100 (Fe) after different immersion times. Each pattern presents the ratios of the changing relative intensity of the a:b = (022):(357) planes, demonstrating the progressive increase in the relative intensity of the (022) plane.

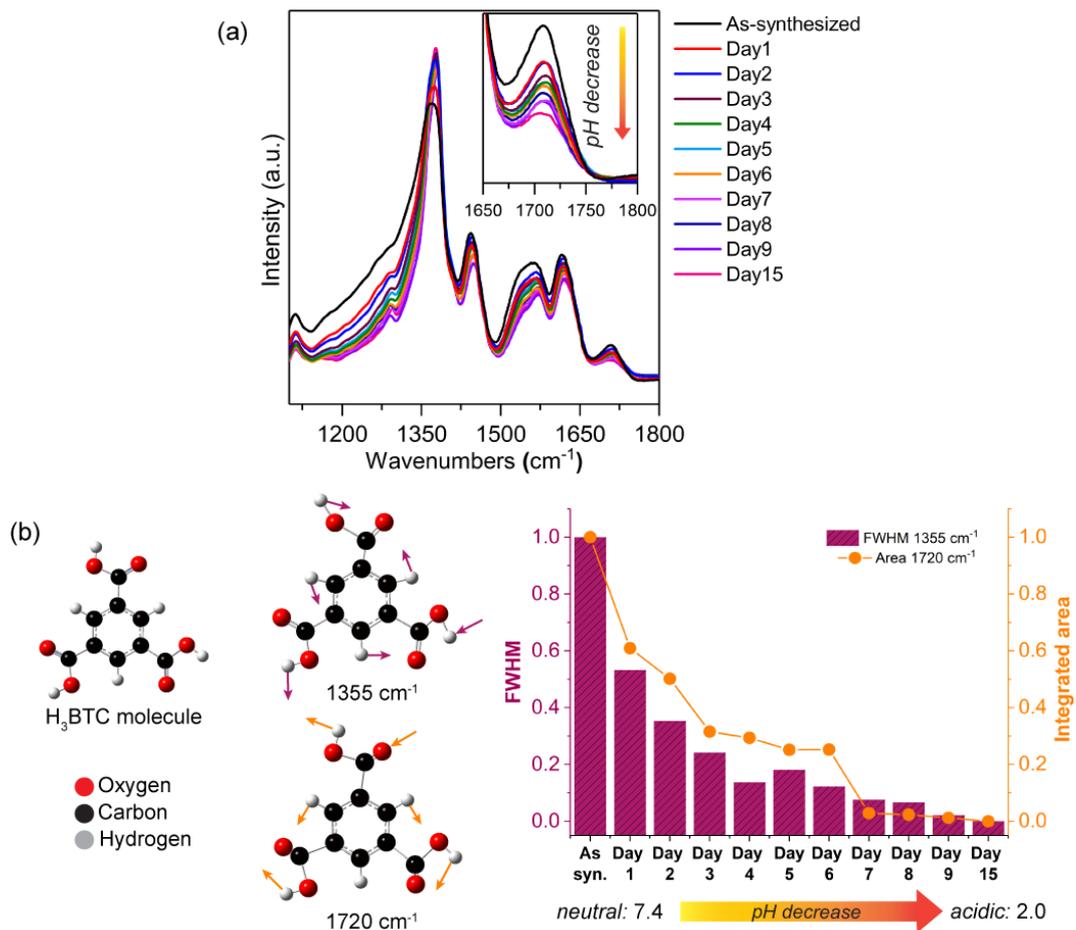

**Figure S2.** (a) Sharpening of the carboxylate vibrational band at ~1355 cm$^{-1}$ and the decrease in intensity of the C=O stretching of H$_3$BTC at ~1720 cm$^{-1}$, expressed numerically by (b) FWHM values and integrated area, respectively. Colour code: O in red, C in black, and H in grey.

**3.2. BET surface area of MIL-100 (Fe) samples**

To acquire deeper insights into the reconstruction process, four samples collected at different immersion times (*i.e.* 2, 4, 7 and 12 days) were selected for nitrogen adsorption and desorption measurements. Evaluation of the nitrogen isotherms shapes (Figure 2b) coupled with the Brunauer–Emmett–Teller (BET) surface area determination (Table S2) allowed us to classify the samples porosity and assess its evolution during the reconstruction process. The change in the pore structure was accompanied by the increase in the BET surface area.

**Table S2.** Comparison of the BET surface area of MIL-100 samples produced *via* various methods

| | Sample | Synthesis method | BET surface area ($m^2.g^{-1}$) | Reference |
|---|---|---|---|---|
| | MIL-100 as-synthesized | Mechanochemistry (manual grinding) | 157 | This work |
| Reconstruction process | MIL-100 Day 2 | Mechanochemistry followed by water immersion | 667 | |
| | MIL-100 Day 4 | | 720 | |
| | MIL-100 Day 7 | | 735 | |
| | MIL-100 Day 12 | | 997 | |
| | 5-FU@MIL-100_REC | REC: encapsulation during reconstruction | 435 | |
| | CAF@MIL-100_REC | | 223 | |
| | ASP@MIL-100_REC | | 225 | |
| Time degraded sample | MIL-100 aged (1.5 years of shelf life) | Mechanochemistry (manual grinding) | 521 | |
| | MIL-100 aged reconstructed | Water immersion | 767 | |
| Mechanically amorphized sample | MIL-100 crushed pellet (10 ton) | Mechanochemistry (manual grinding) | 14 | |
| | MIL-100 crushed pellet reconstructed | Water immersion | 276 | |
| MIL-100 (Fe) | | Mechanochemistry (high pressure and temperature) | 1940 | Han *et al.*[7] |
| | | Mechanochemistry (Liquid assisted grinding - ball mill) | 1033 | Pilloni *et al.*[8] |
| | | Mechanochemical (kitchen grinder) | 255 | Samal *et al.*[9] |
| | | Solvothermal (high temperature) | 1836 | Zhang *et al.*[10] |
| | | Solvothermal (high pressure and temperature) | 1750 | Chen *et al.*[11] |
| | | Solvothermal (high pressure and temperature) | 1223.32 | Zhang *et al.*[12] |
| | | Solid state synthesis (high pressure and temperature) | 110.49 | Chaturvedi *et al.*[13] |

We have compared the synthesis parameters (*i.e.* total synthesis time and required temperature) of approaches conventionally used for the fabrication of MIL-100 (Fe), some newly employed mechanochemical methods and the mechanochemical-reconstruction method herein applied based on the resulting BET surface area achieved (Figure S3).

Our as-synthesized MIL-100 (Fe) material has shown higher surface area when compared to the material obtained *via* the solid-state approach (Chaturvedi *et al.*), conducted under high pressure conditions. The surface area of our material is significantly improved after the reconstruction process (*i.e.* MIL-100 Day 12) and is comparable to those obtained by solvothermal and ball milling approaches (Pilloni *et at.*), which require the use of more extreme conditions (high temperature and pressure) and the use of corrosive toxic agents.

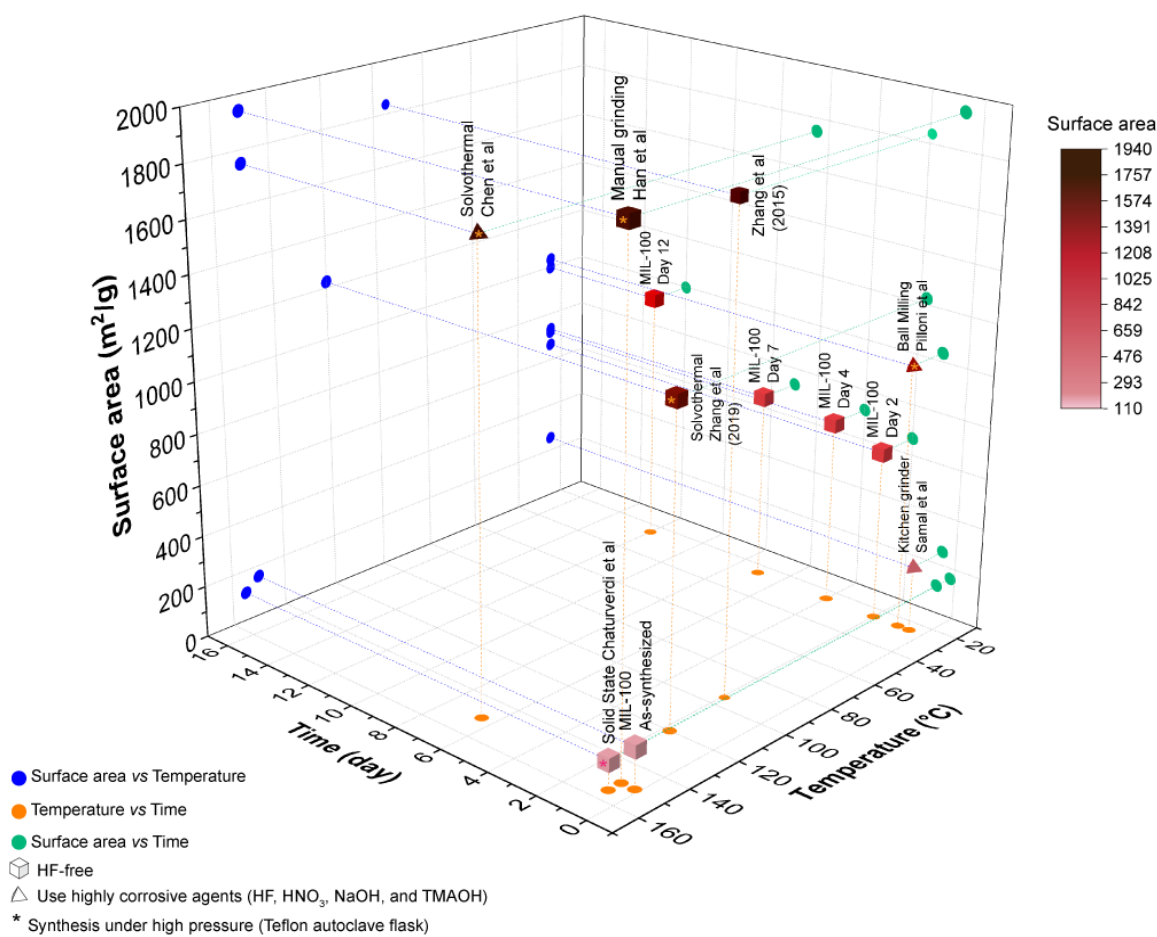

**Figure S3.** Comparison of BET surface area, synthesis temperature and total synthesis time of MIL-100 (Fe) samples prepared in this work *via* the reconstruction technique and the reported BET surface area of MIL-100 (Fe) fabricated *via* various synthetic methods.
S8

### 3.3. Morphological characterization of MIL-100 (Fe) sample

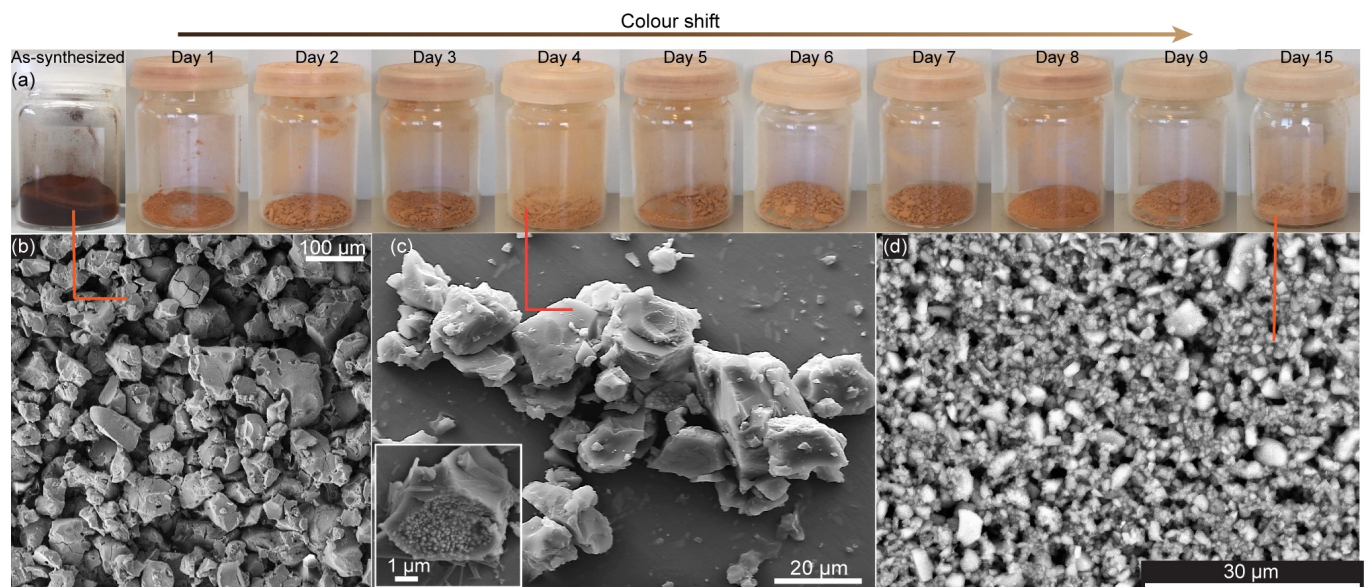

**Figure S4.** SEM images of MIL-100 (Fe) samples. (a) Photographic representation of MIL-100 (Fe) samples collected at different immersion time intervals. The change in the sample color can be associated to changes in the aggregation of particles as observed in (b), (c) and (d), respectively.

MIL-100 (Fe) Day 12 sample was used for the fabrication of pellets under different uniaxial forces. All pellets were produced from the same batch of material, and presented good mechanical strength, being able to be manipulated by hand. As the pelletizing force increased, a colour shift was observed with the darkening of the pellets as a result of the denser packing of MIL-100 (Fe) nanoparticles. This effect reproduces the optical changes shown in Figure S4.

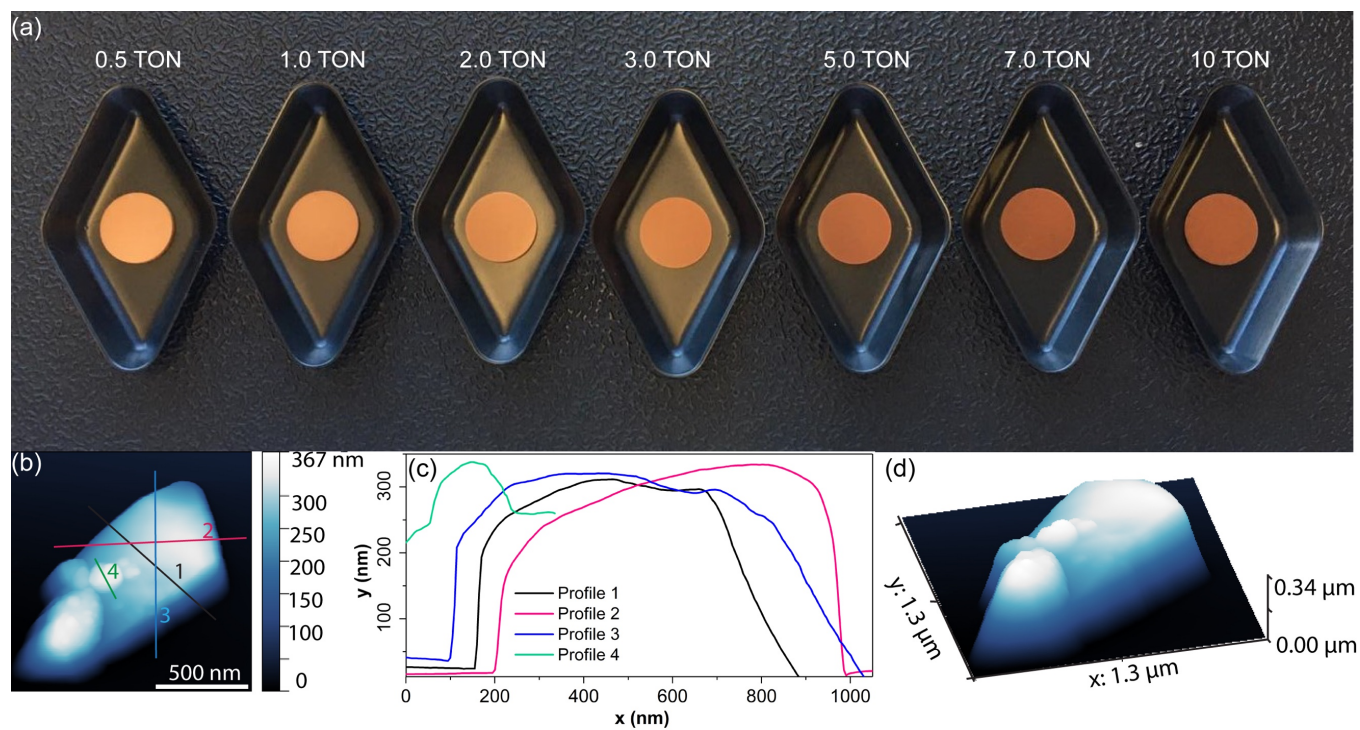

**Figure S5**. (a) Pellets of MIL-100 (Fe) samples produced by the application of various uniaxial forces. The packing of the MIL-100 particles demonstrates that shifts in the colour of the pellets, going from light to dark brown, are observed as the particles become more densely packed (i.e. subjected to 0.5 to 10 ton). (b) AFM image of MIL-100 (Fe) Day 12 crystals used to produce the pellets displayed in (a). (c) Height profile of crystals revealing the size and shape of MIL-100 (Fe) Day 12 crystals, with a rectangular-like shape of hundreds of nanometers. (d) 3D view of crystal displayed in (b).

### 3.4. Reconstruction of aged-MIL-100 (Fe)

Using the reconstruction strategy, we were able to recover the crystallinity of MIL-100 (Fe) samples with a shelf age of 1.5 years. Throughout this time, the samples have been stored in a standard sealed vial. As observed in Figure S6a, the aged-MIL-100 (Fe) showed signs of decomposition/collapse due to the change in the relative intensity of the planes (022):(357), going from 1:025 to 1:0.99 for the as-synthesized and aged samples, respectively. The (022) plane seemed to suffer the most from the aging of the sample, with large reduction of intensity. As it can be noted from Figure S6c, the (022) peak has shown large broadening after the aging process.

*Via* the reconstruction process (Figure S6b) the crystallinity of the material was increasingly recovered with the increase of immersion time. Not only the relative intensity of the samples presented a massive change (from 1:0.99 to 1:0.26), but also the FWHM values of the (022) peak were fully restored.

Concomitantly, as shown in Figure S7, the broadening of the carboxylate vibration band at 1355 cm$^{-1}$ in the aged sample and the sharpening of this same band after the reconstruction process indicates the increase in the structural symmetry of the sample. Moreover, the BET surface area presented a recovery of 96.7% against the initial value, demonstrating the reconstruction of the materials porosity.

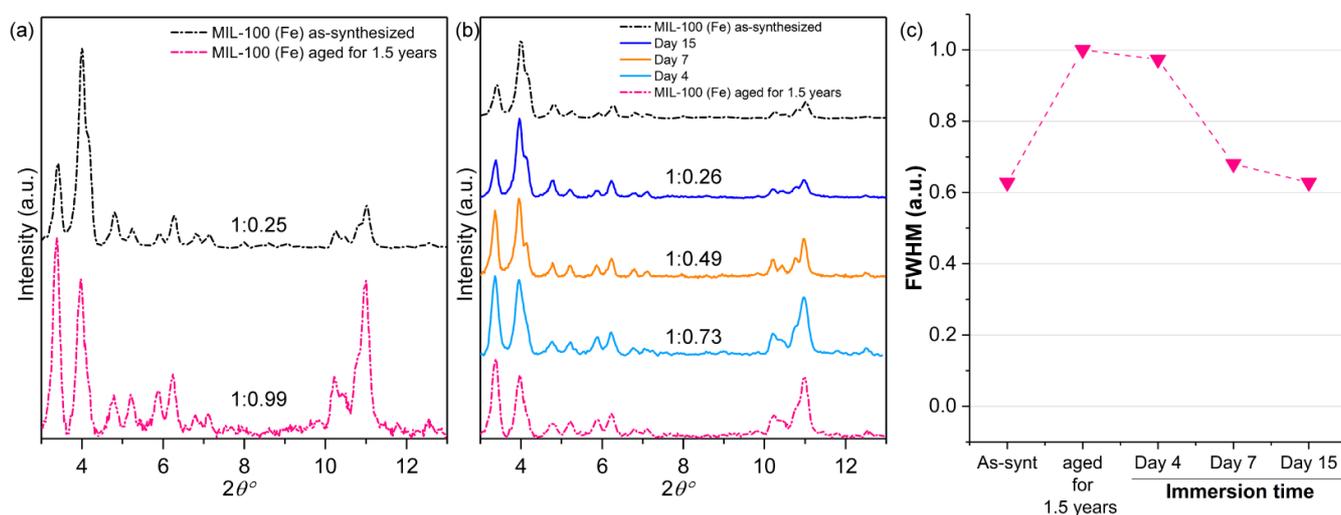

**Figure S6.** PXRD patterns of aged samples used in the reconstruction process. (a) PXRD pattern of fresh MIL-100 (Fe) (black trace) and the aged sample (pink trace). (b) PXRD patterns showing the recovery of crystallinity of the aged sample after immersion in DI water. (c) FWHM of the (022) plane.

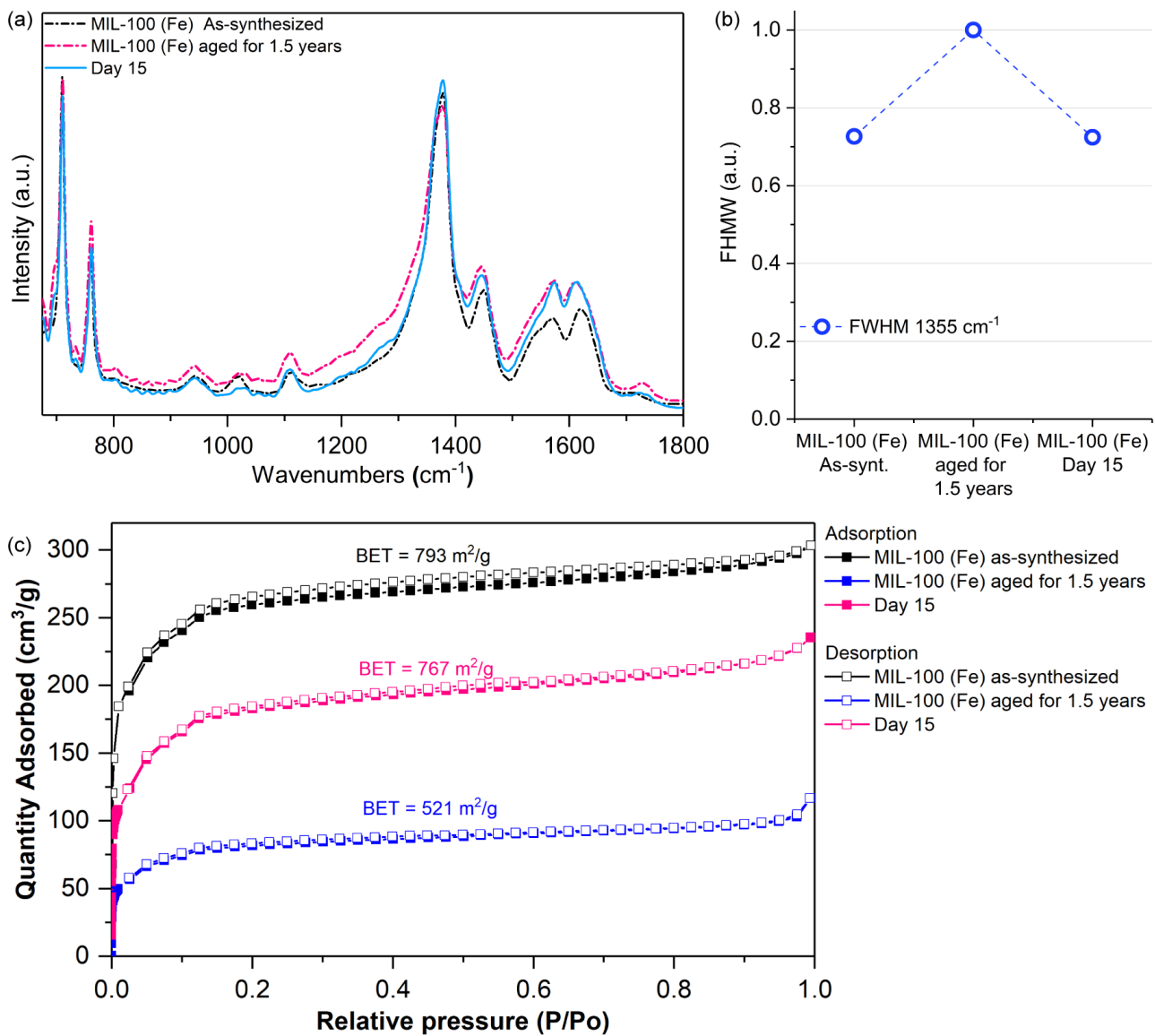

**Figure S7.** Analysis of vibrational data of the reconstructed aged sample. (a) ATR-FTIR spectra of aged samples under reconstruction. (b) FWHM values of vibration band at ~1355 cm$^{-1}$ highlighting the sharpening of carboxylate vibration after the reconstruction process. (c) Nitrogen adsorption and desorption isotherms of time degraded MIL-100 (Fe) samples under reconstruction. Samples were activated at 150 °C under high vacuum for 12 hours prior to the N$_2$ adsorption measurements at 77 K. The BET surface area presented a recovery of 96.7% versus the initial value.

The reconstruction process has also been successfully used for the recovery of crystallinity of mechanically amorphized samples. As demonstrated in Figure S8, the diffraction peaks at 4° and 11° vanish after the pelletizing process. Concomitantly, as it can be seen from Figure S9c, the material is almost non-porous with a much reduced surface area of 14.22 m$^2$/g.

After relaxation of the structure (*i.e.* crushing the pellet back to powder) some of the most intense diffraction peaks reappeared. However, the relative intensity of (022):(357) peaks is very distinct from the original powder used for the production of the pellets. The above ratio was almost completely restored after 7 days of immersion in DI water.

Similar to what was observed during the reconstruction of the aged samples, Figure S9 shows the broadening of the carboxylate vibration band at 1355 cm$^{-1}$ after pelletizing and the sharpening of the same band after the reconstruction process, indicating the increase in the structural symmetry of the sample.

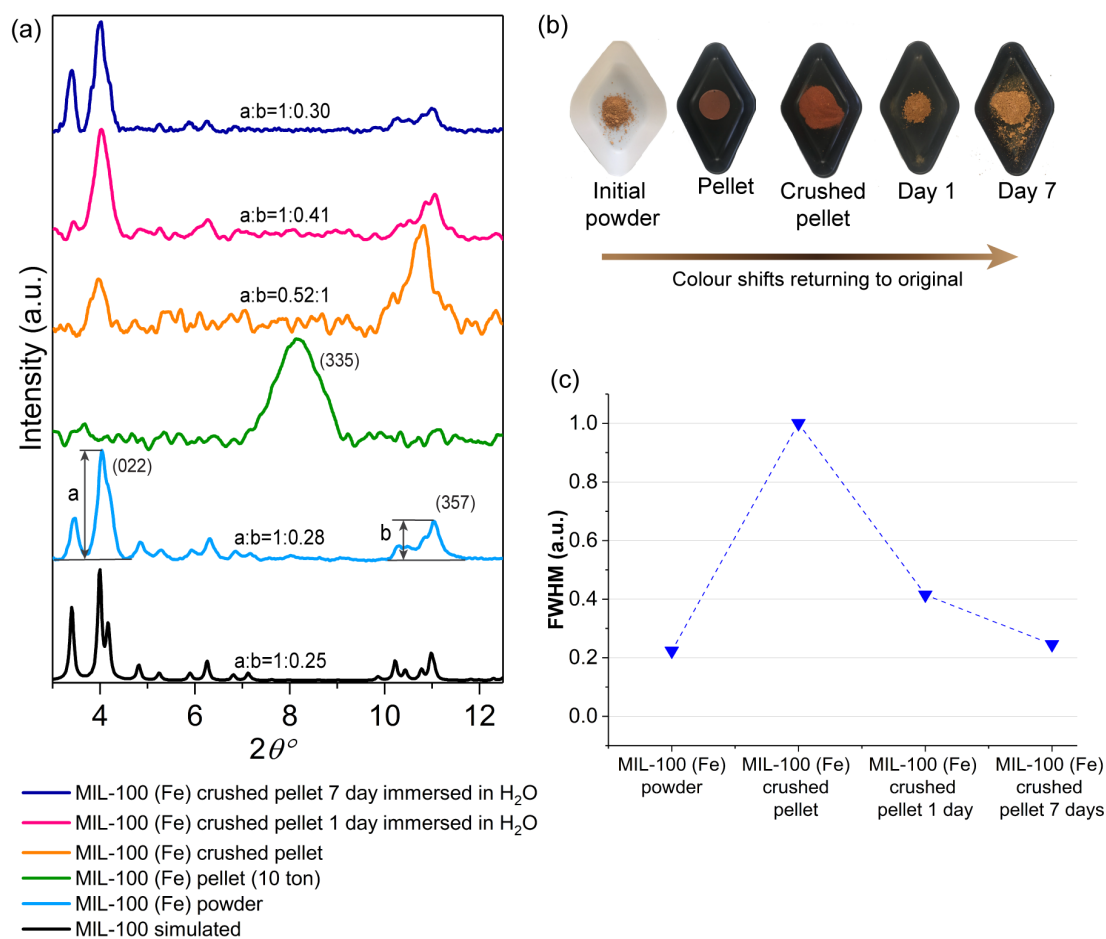

**Figure S8.** Reconstruction of mechanically amorphized samples. (a) PXRD patterns showing the progressive change in the relative intensity of the diffraction peaks. (b) Colour shift presented by the samples during the reconstruction process in which the original colour is restored demonstrating changes in the microstructure/optical properties of the material. (c) FWHM used to assess the sharpening of the (022) peak. As it can be observed, after the pelletizing, the diffraction peak corresponding to the (022) plane becomes very broad. However, the FWHM values are virtually restored to the pristine material values after the reconstruction process.

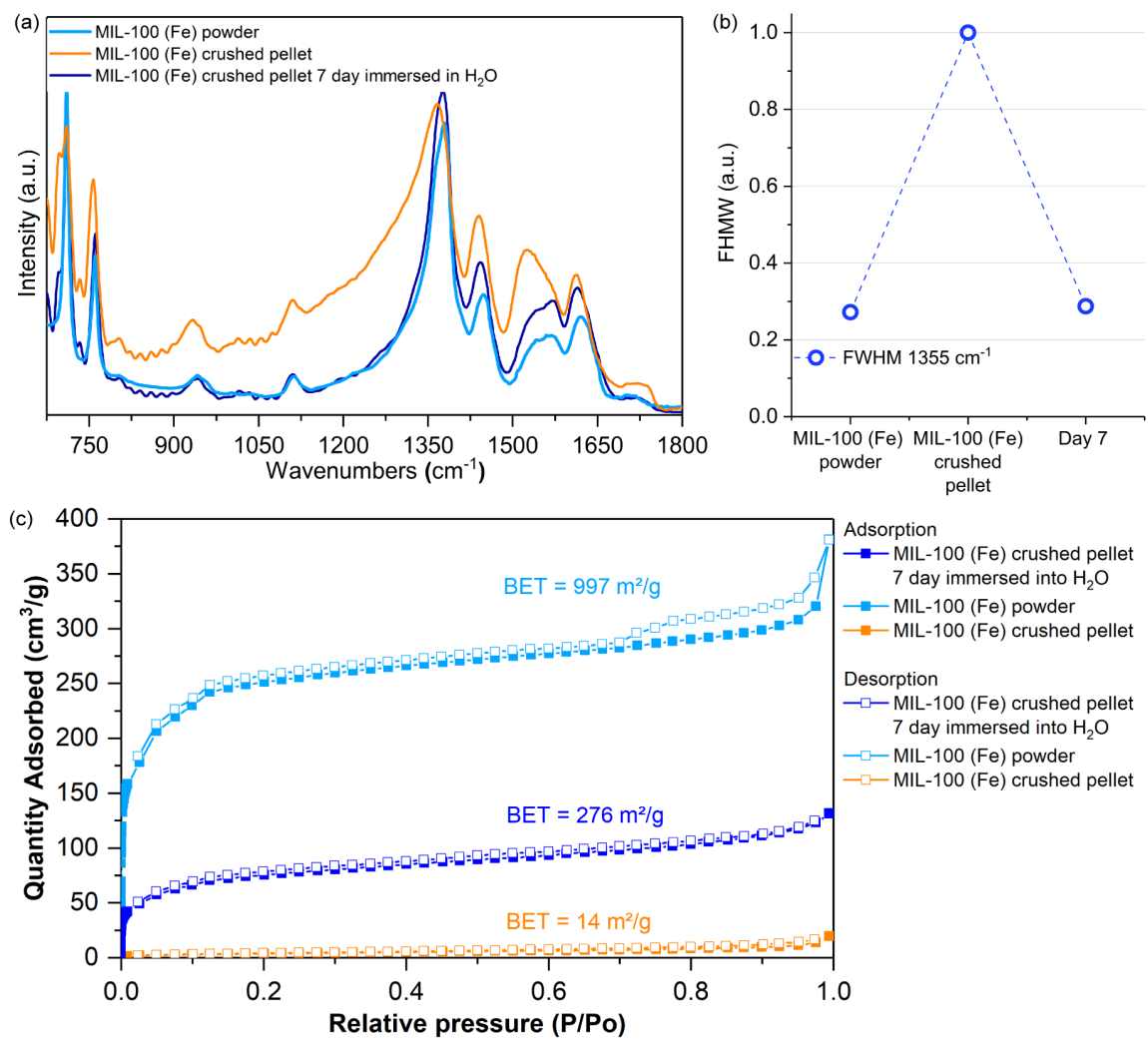

**Figure S9.** Analysis of vibrational data of reconstruction of pellet sample. (a) ATR-FTIR spectra of pelletized samples under reconstruction. (b) FWHM values of vibrational band at ~1355 cm$^{-1}$ highlighting the sharpening of carboxylate vibration after the reconstruction process. (c) Nitrogen adsorption and desorption isotherms of mechanically amorphized MIL-100 (Fe) samples under reconstruction. Samples were activated at 150 °C under high vacuum for 12 hours prior to the N$_2$ adsorption measurements at 77 K. The BET surface area presented a recovery of only 27.7% compared to its initial value.

### 3.5. Examination of organic ligand H₃BTC INS spectrum

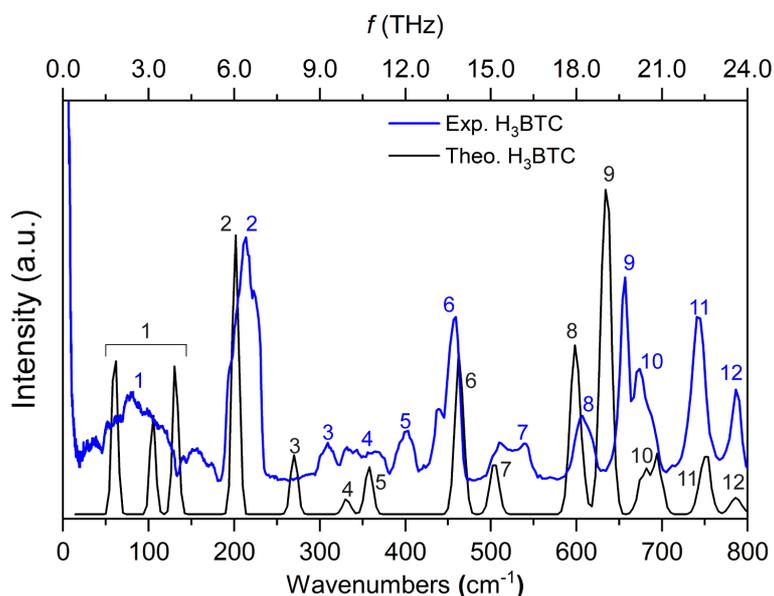

**Figure S10.** Comparison between experimental (blue) and theoretical (black) DFT spectra of H₃BTC. Numbers were used to facilitate the identification of the vibrational modes. Good agreement between both spectra is observed in the low energy region.

**Table S3.** Description of vibrational modes of H₃BTC in the range of 0-800 cm$^{-1}$ [14]

| Mode no. | Theo. (cm$^{-1}$) | Exp. (cm$^{-1}$) | Description |
|---|---|---|---|
| 1 | 20-133 | 21-183 | γCOOH, γCO, and γ ring including trampoline-like motion |
| 2 | 218 | 202 | γCO |
| 3 | 272 | 273 | δCO |
| 4 | 332 | 334 | δCO |
| 5 | 361 | 358 | γ ring and γOH |
| 6 | 470 | 464 | δCC and δOH |
| 7 | 505 | 506 | δCO and δCH |
| 8 | 603 | 600 | γ ring and γOH |
| 9 | 635 | 636 | δOH and δCH |
| 10 | 674 | 685 | |
| 11 | 714 | 750 | δCO and δCC |
| 12 | 780 | 786 | |

δ, in-plane bending, γ, out-of-plane bending.

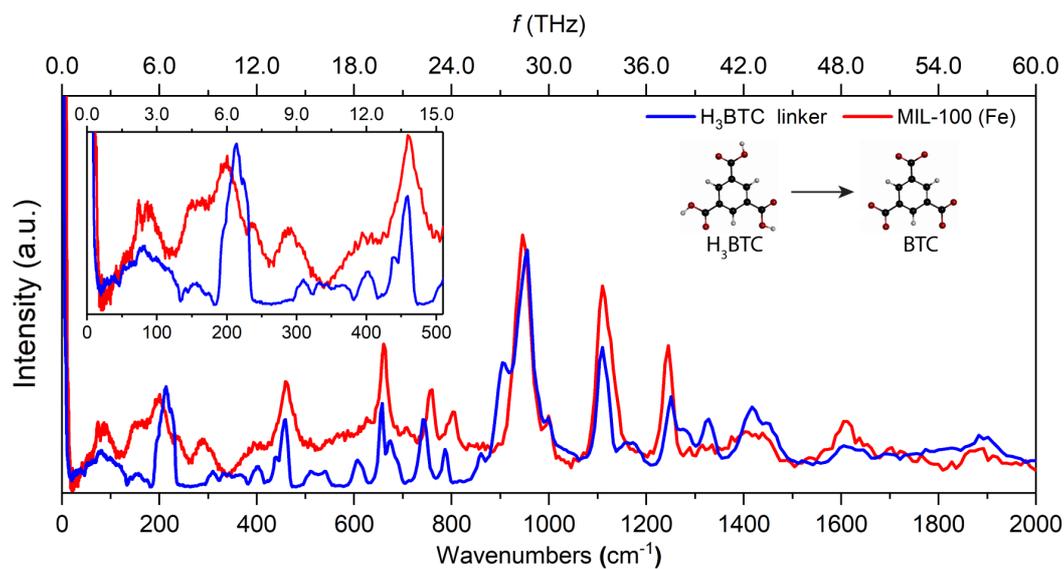

**Figure S11.** INS spectra of pristine MIL-100 (Fe) sample (red) and organic linker $H_3BTC$ (blue). Good agreement is observed between framework modes and organic linker modes, helping us to pinpoint in the MOF spectrum the vibrational modes related to the organic linker. Note that the $H_3BTC$ spectrum was scaled down by a factor of 0.3 to facilitate comparison with the MOF spectrum. Colour code: O in red, C in black, and H in grey.

## 3.6. Thermal stability of MIL-100 (Fe) reconstructed samples

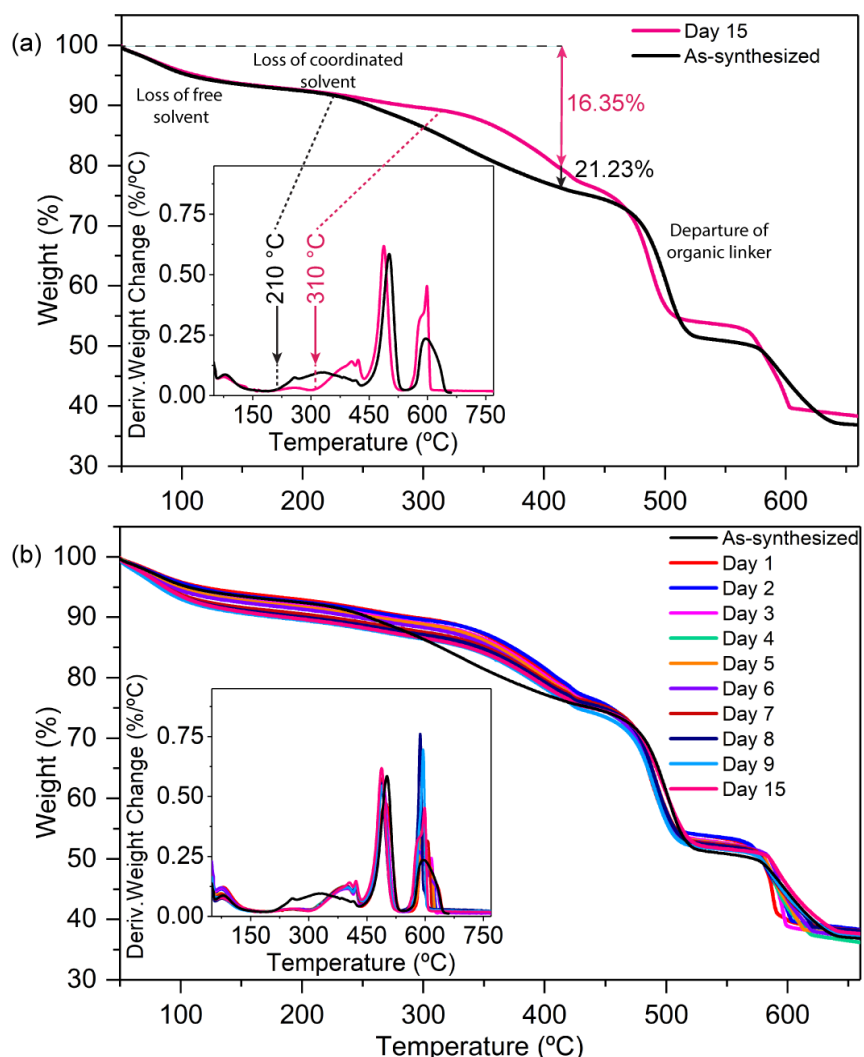

**Figure S12.** TGA of MIL-100 (Fe) samples with insets showing the differential material decomposition behaviour as a function of temperature. (a) Contrast between the as-synthesized MIL-100 (Fe) and reconstructed MIL-100 (Fe) after 15 days of immersion time. An increase in the thermal stability of the material is observed (~100 °C difference in the initial decompostion temperature of the material). (b) Contrast between TGA of as-synthesized sample and samples subjected to different immersion times.

Table S4 displays details of the decomposition process of MIL-100 (Fe) sample acquired from the TGA plots (Figure S12). A large increase in the initial decomposition temperature of the crystalline MIL-100 (Fe) in comparison to the as-synthesized material can be observed. A decrease (~20%) in the rate of the composition between 210-450 °C was also noticed. Meanwhile, the maximum rate of decomposition, the temperature in which half of the material has decomposed and the final residue have remained virtually unchanged.

**Table S4.** Analysis of decomposition of MIL-100 (Fe) samples.

|  | Initial decomposition temperature | Rate of decomposition (210 °C - 450 °C) | Maximum rate of decomposition | Temperature of half decomposition | Final residue (at 660 °C) |
| --- | --- | --- | --- | --- | --- |
| As-synthesized | 210 °C | -0.076 %/°C | -0.275 %/°C | 574 °C | 37% |
| MIL-100 (Fe) Day 15 | 310 °C | -0.061 %/°C | -0.289 %/°C | 578 °C | 38% |

### 3.7. SEM images of drug@MIL-100 systems

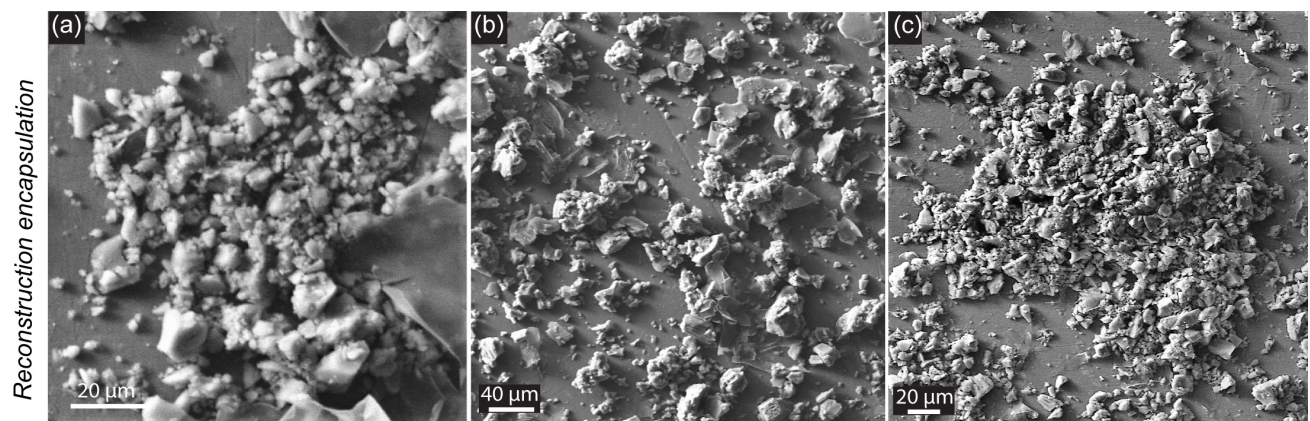

**Figure S13.** SEM images of guest@MIL-100 systems. (a) 5-FU@MIL-100_REC, (b) CAF@MIL-100_REC, and (c) ASP@MIL-100_REC.

### 3.8. Diffraction data analysis of drug@MIL-100 systems

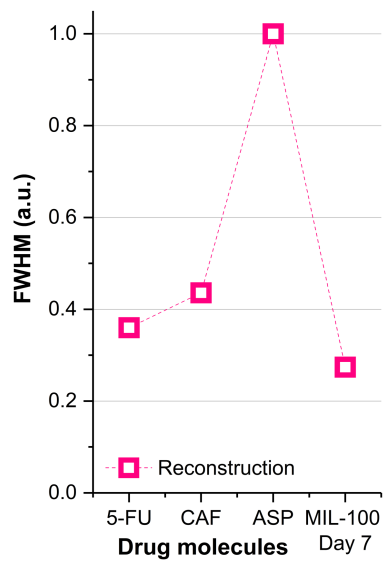

**Figure S14.** FWHM used to assess the sharpening of the (022) peak in the drug@MIL-100 systems, contrasting the effect of the different guest drug molecules being used. The plots evidence the strong effect on the material crystallinity that aspirin has in comparison to the other drug molecules when the reconstruction strategy is applied for the encapsulation of the guest molecules. To facilitate the comparison, the values have been normalized in relation to the highest FWHM value among this set of samples.

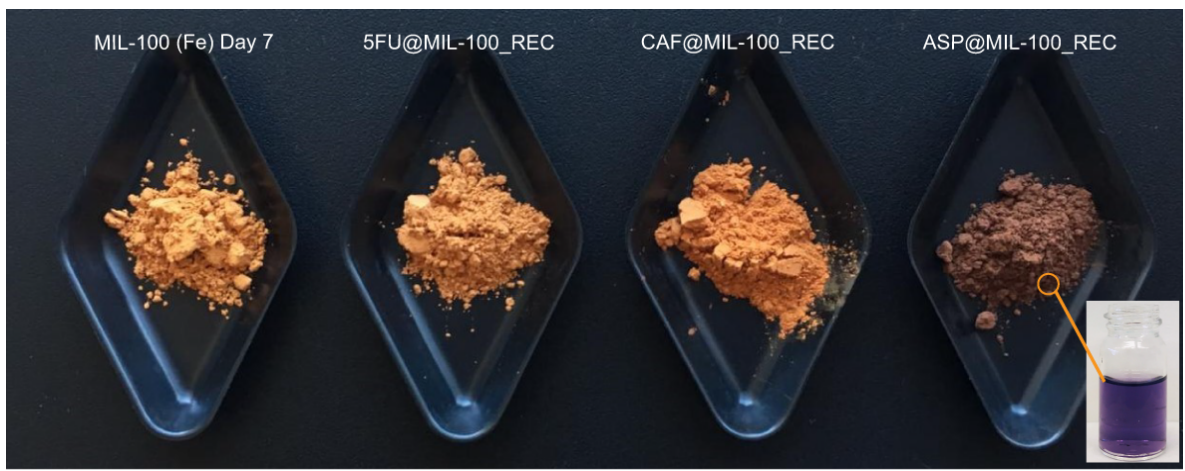

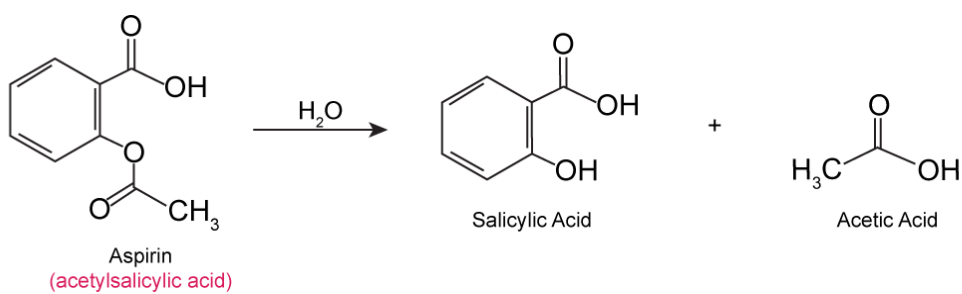

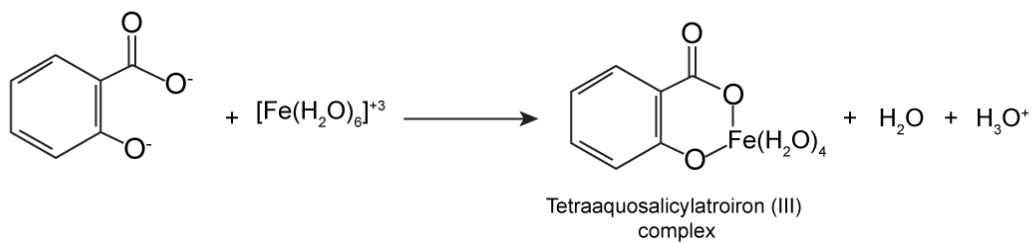

**Figure S15.** Photograph showing the various MIL-100 (Fe) samples after the encapsulation-reconstruction process, showing a distinct colour difference between 5-FU@MIL-100_REC and CAF@MIL-100_REC to ASP@MIL-100_REC. This is due to the formation of a violet aspirin-iron complex known as tetraaquosalicylatroiron (III) complex (inset). When in contact with moisture, aspirin dissociates into acetic and salicylic acid. The latter then can react with acidified iron (III) ions to form the violet complex.

### 3.9. INS spectra of MIL-100 (Fe) and guest@MIL-100 systems

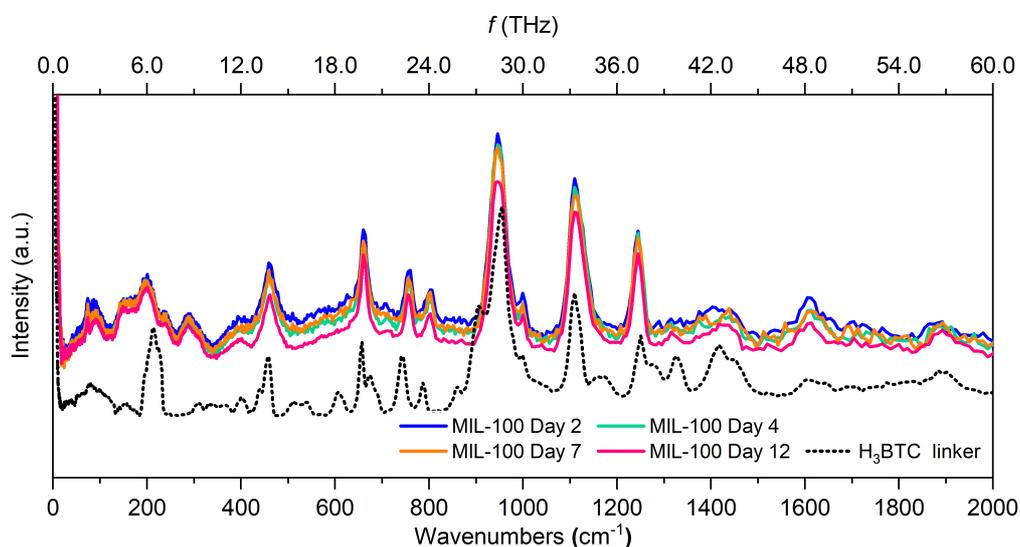

**Figure S16.** INS spectra of MIL-100 (Fe) samples collected after different immersion times. In black is the spectrum of H₃BTC linker which was scaled down by a factor of 0.2 to facilitate the comparison with MOF spectra.

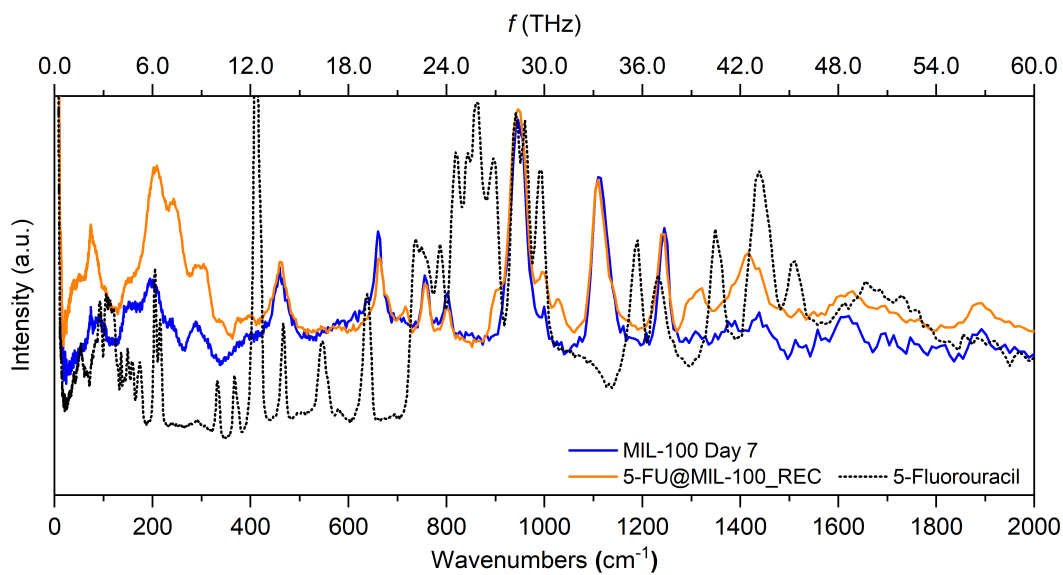

**Figure S17.** INS spectra of 5-FU@MIL-100 yielded by the reconstruction and *in situ* encapsulation techniques. Spectrum of 5-FU presented in black was scaled down by a factor of 0.4 to facilitate the comparison with the guest@MIL-100 systems.

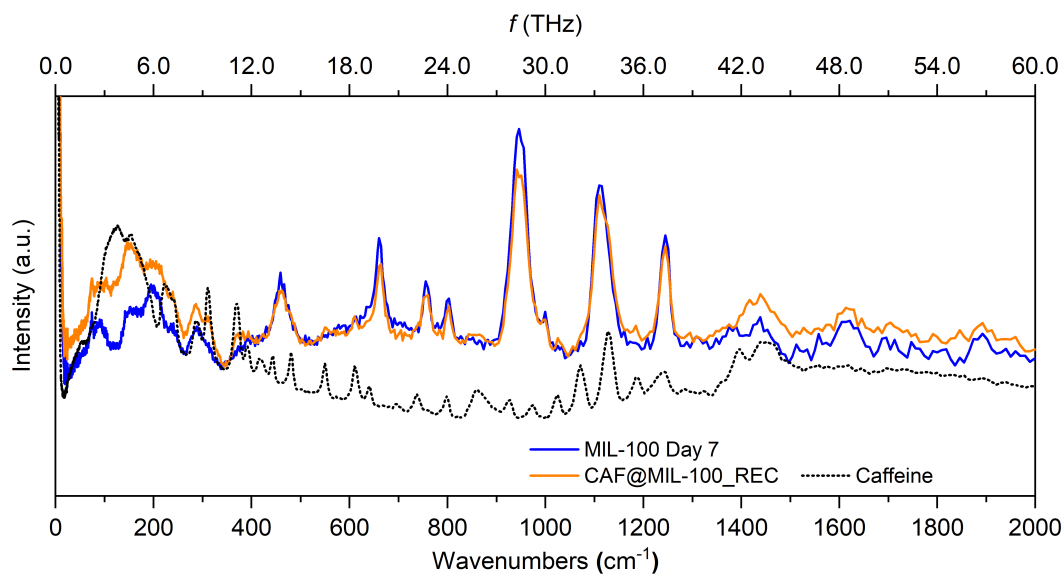

**Figure S18.** INS spectra of CAF@MIL-100 yielded by the reconstruction and *in situ* encapsulation techniques. Spectrum of caffeine presented in black was scaled down by a factor of 0.2 to facilitate the comparison with the guest@MIL-100 systems.

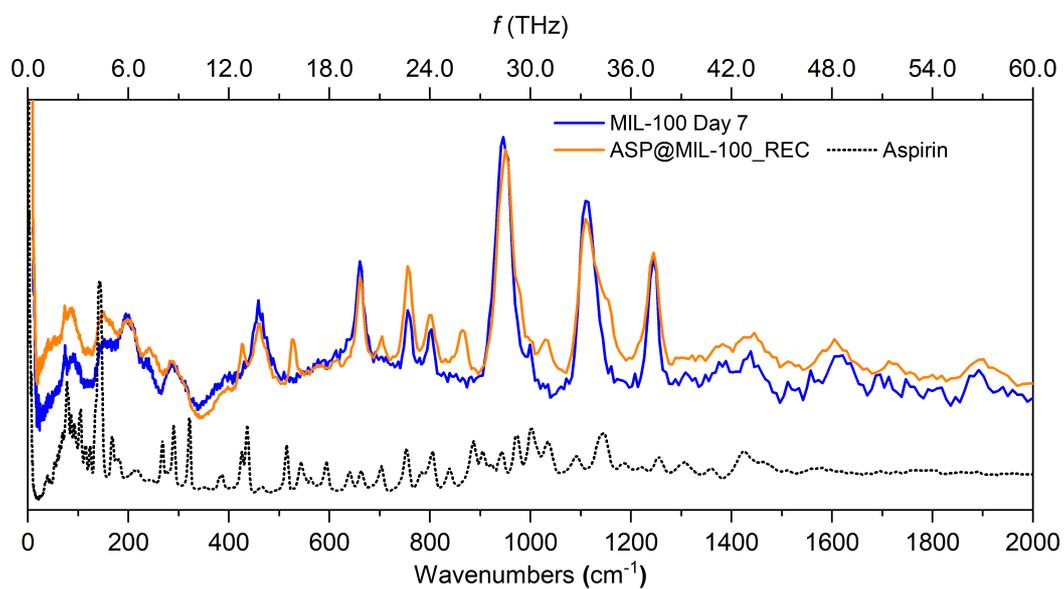

**Figure S19.** INS spectra of ASP@MIL-100 yielded by the reconstruction and *in situ* encapsulation techniques. Spectrum of aspirin presented in black was scaled down by a factor of 0.1 to facilitate the comparison with the guest@MIL-100 systems.

## 3.10. ATR-FTIR spectra of guest@MIL-100 systems

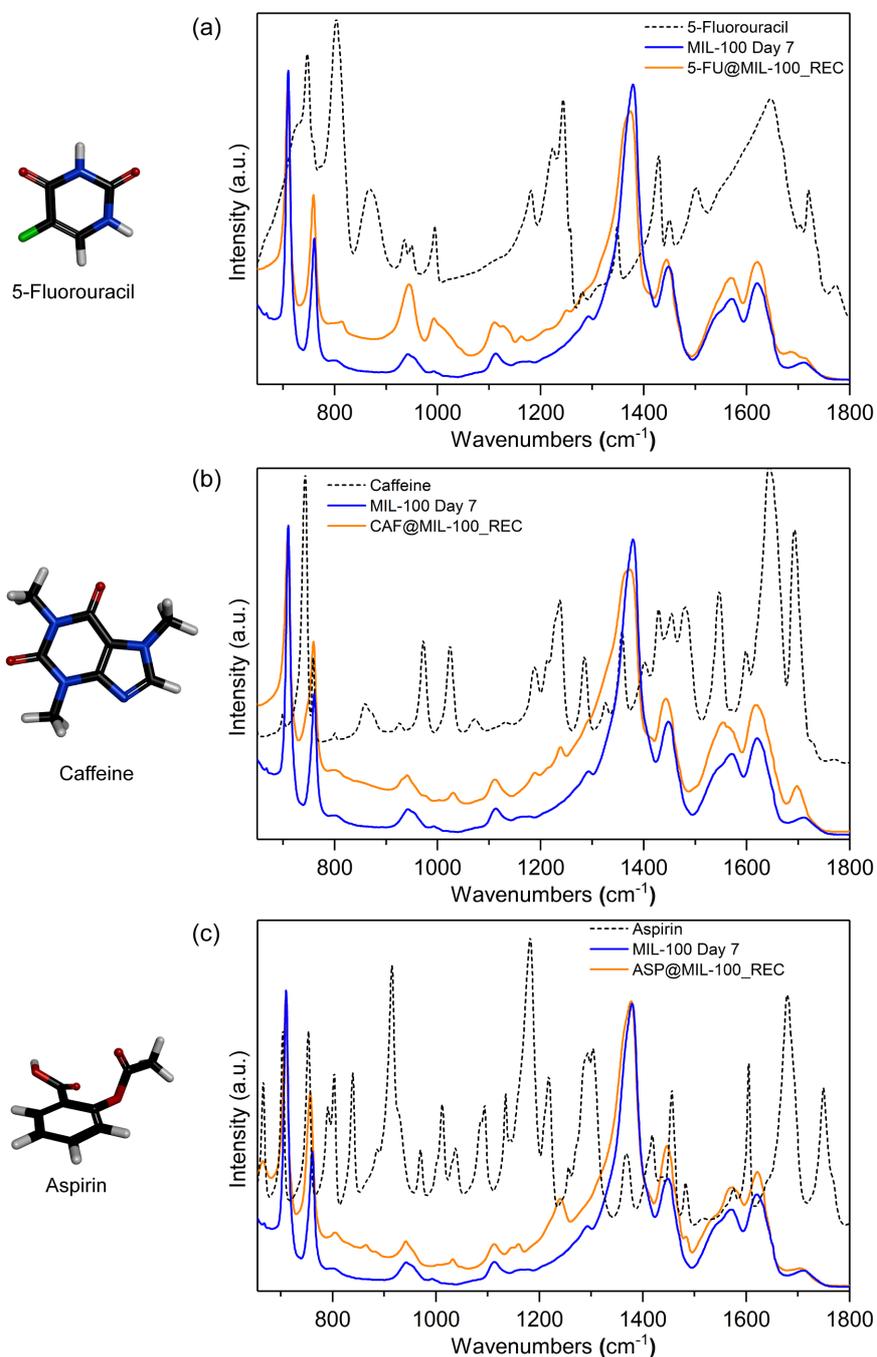

**Figure S20.** ATR-FTIR spectra of guest@host systems displaying (a) 5-FU@MIL-100_REC, (b) CAF@MIL-100_REC, and (c) ASP@MIL-100_REC. For comparison, the spectrum of MIL-100 (Fe) after 7 days of immersion time was also presented. Colour code: O in red, C in black, H in grey, N in navy blue, F in green.

**3.11. Evaluation of guest encapsulation *via* thermogravimetric analysis and nitrogen adsorption/desorption**

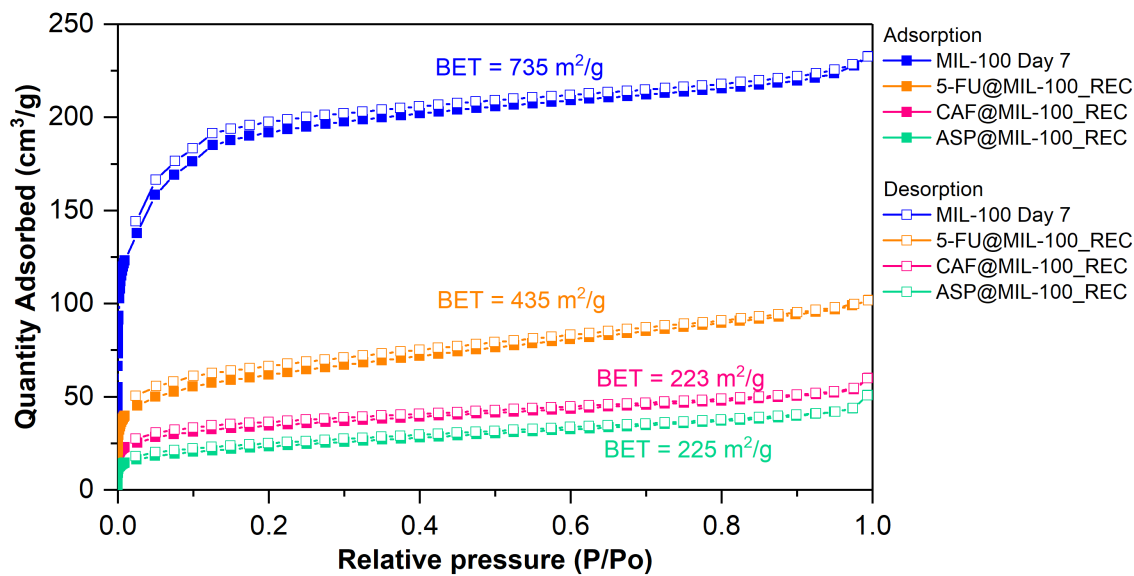

**Figure S21.** Nitrogen adsorption and desorption isotherms of 5-FU@MIL-100_REC, CAF@MIL-100_REC, and ASP@MIL-100_REC systems.

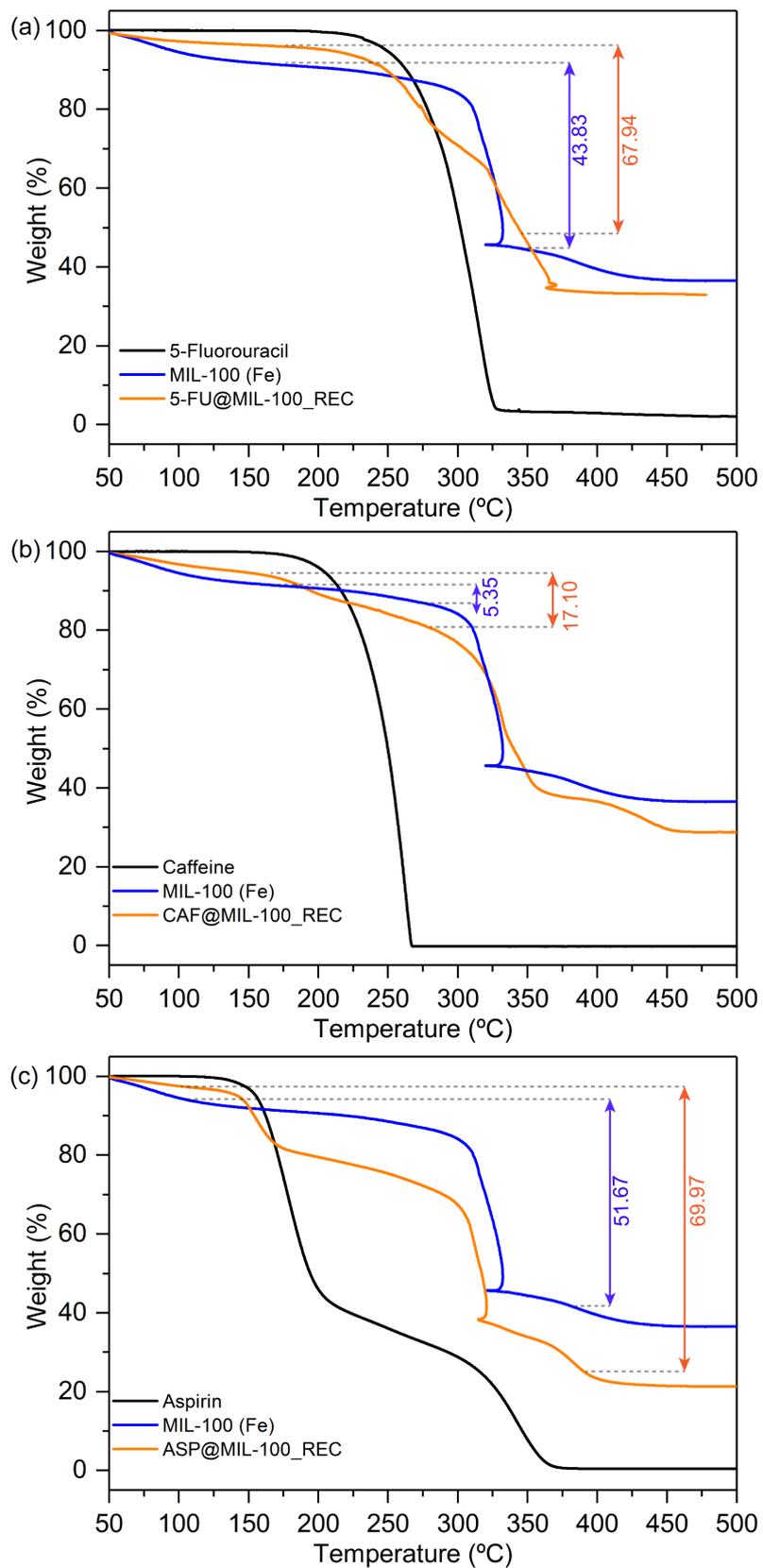

**Figure S22.** TGA of drug@MIL-100 systems showing the material decomposition behaviour as a function of temperature. (a) 5-FU@MIL-100_REC, (b) CAF@MIL-100_REC, and (c) ASP@MIL-100_REC.

**Table S5.** Comparison of drug loading of drug@MIL-100 samples produced *via* various methods

| | Sample | Synthesis method | Drug loading (wt. %) | Drug loading (g/g of MOF) | Reference |
|---|---|---|---|---|---|
| 5-Fluorouracil | 5-FU@MIL-100_REC | Reconstruction encapsulation | 35.5 wt.% | 0.6 | This work |
| | 5-FU@MIL-100 | Immersion into drug solution | 66.0 wt.% | - | Thi *et al.* [15] |
| | 5-FU@MIL-88 | | 28.0 wt.% | - | |
| | 5-FU@MIL-53 | | 13.1 wt.% | - | |
| Caffeine | CAF@MIL-100_REC | Reconstruction encapsulation | 64.7 wt.% | 1.8 | This work |
| | CAF@MIL-100 | Immersion into drug solution | 49.7 wt.% | - | Cunha *et al.* [16] |
| | | | 24.2 wt.% | - | Horcajada *et al.* [17] |
| | | | 52.4 wt.% | - | Márquez *et al.* [18] |
| Aspirin | ASP@MIL-100_REC | Reconstruction encapsulation | 70.0 wt.% | 2.3 | This work |
| | ASP@MIL-100 | Immersion into drug solution | 24.8 wt.% | - | Rojas *et al.* [19] |
| | | | - | 1.8 | Singco *et al.* [20] |
| | ASP@MIL-127 | | - | 0.14 | Rojas *et al.* [21] |

The drug loading was calculated from the TGA plots analysis *via* the formula:

$$\text{wt.}\% = \frac{m_{\text{loss (drug@MOF)}} - m_{\text{loss (MOF)}}}{m_{\text{loss (drug@MOF)}}}$$

where $m_{\text{loss (drug@MOF)}}$ is the weight loss of the drug@MOF systems and $m_{\text{loss (MOF)}}$ is the weight loss of the host MOF.

### 3.12. 5-Fluorouracil DFT calculations details

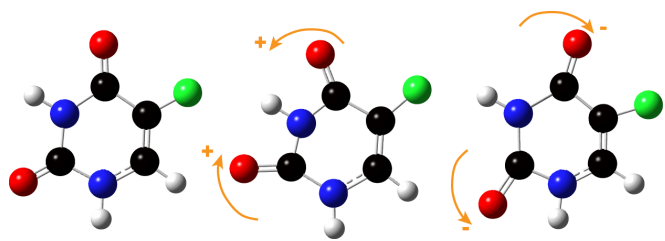

**Figure S23.** Illustration of the in-plane bending of OCNCO at ~12.3 THz (~410 cm$^{-1}$). Colour code: O in red, C in black, H in grey, N in navy blue, F in green.

### 3.13. Guest and host size comparison

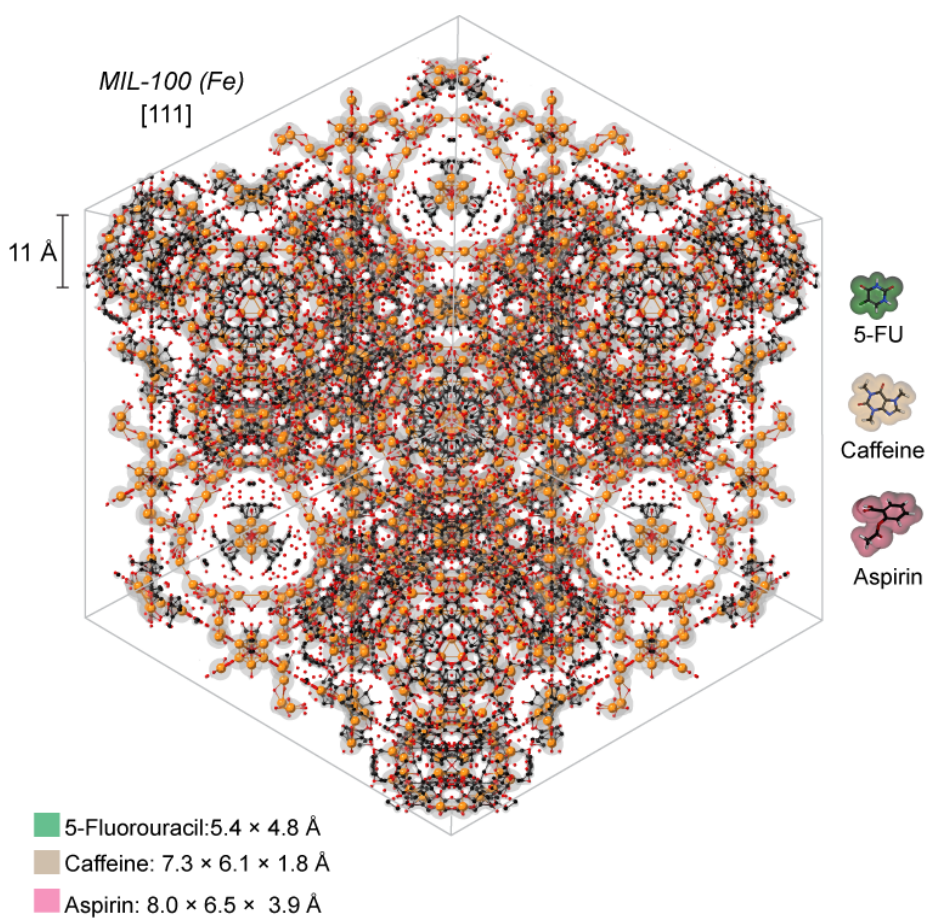

**Figure S24.** Schematic representation of host and guest molecular sizes. The comparison showcases how the mesocages of MIL-100 (Fe) can accommodate multiple of the chosen drug molecules within the MOF pores.